\documentclass{IEEEtaes}

\usepackage{color,array,amsthm}
\usepackage{graphicx}

\usepackage{algorithm}      
\usepackage{algpseudocode}  
\usepackage{comment}

\usepackage{paralist}
\usepackage{amssymb}
\usepackage{amsmath}
\usepackage{subcaption}
\usepackage{caption}
\renewcommand{\thefigure}{\arabic{figure}} 

\usepackage{booktabs}     
\usepackage{multirow} 

\usepackage{stfloats}				
\usepackage{float}                  

\usepackage[nocompress]{cite}
\usepackage{orcidlink}
\usepackage{hyperref}

\jvol{XX}
\jnum{XX}
\jmonth{XXXXX}
\paper{1234567}
\pubyear{2026}
\doiinfo{TAES.2026.Doi Number}

\begin{document}

	\title{Predictive Beam Tracking for Sea-Air Communication under Wave-Induced Disturbances} 
	
	\author{RUI ZHANG~\orcidlink{0009-0006-6793-8310}}	
	\author{FUWANG DONG~\orcidlink{0000-0001-6522-9082}}
	\member{Member, IEEE}
		\author{WEI WANG~\orcidlink{0000-0002-4392-1884}}
	\member{Senior Member, IEEE}
		\affil{Harbin Engineering University, Harbin, China}
		\author{ZHEN DU~\orcidlink{0000-0002-5882-1595}}
	\member{Member, IEEE}
	\affil{Nanjing University of Information Science and Technology, Nanjing, China}
	
	\receiveddate{Manuscript received XXXXX 00, 0000; revised XXXXX 00, 0000; accepted XXXXX 00, 0000.\\
		This work was supported in part by the National Natural Science Foundation of China under Grant No. 62271163, in part by the Fundamental Research Funds for the Central Universities under Grant No. 3072025CFJ0409, and in part by the Heilongjiang Province Postdoctoral Research Funds under Grant No. 3236330159. \itshape (Corresponding author: Fuwang Dong.)}
		
	\authoraddress{
		Rui Zhang, Fuwang Dong, and Wei Wang are with the College of Intelligent Systems Science and Engineering, Harbin Engineering University, Harbin 150001, China (e-mail: \href{mailto:azhangrui407@hrbeu.edu.cn}{azhangrui407@hrbeu.edu.cn}; \href{mailto:dongfuwang@hrbeu.edu.cn}{dongfuwang@hrbeu.edu.cn}; \href{mailto:wangwei407@hrbeu.edu.cn}{wangwei407@hrbeu.edu.cn}).
		Zhen Du is with the School of Electronic and Information Engineering, Nanjing University of Information Science and Technology, Nanjing 210044, China (e-mail: \href{mailto:duzhen@nuist.edu.cn}{duzhen@nuist.edu.cn}).
		}
		
	\markboth{RUI ZHANG ET AL.}{
Predictive Beam Tracking for Sea-Air Communication under Wave-Induced Disturbances
		}
	\maketitle
	
\begin{abstract}
In the sea-air communications system consisting of an uncrewed aerial vehicle (UAV) and an uncrewed surface vehicle (USV), the six-degree-of-freedom (6-DoF) motion of the USV under wave-induced disturbances may lead to severe misalignment of communication beams. 
In this paper, we develop a wide- and narrow-beam switching scheme based on an extended Kalman filter (EKF) algorithm. 
Specifically, in the wide-beam mode, the UAV transmits an integrated sensing and communication (ISAC) signal to provide full USV coverage and estimate the position of the communication receiver (CR) mounted on the USV. 
In the narrow-beam mode, the UAV forms a highly directional beam based on the position estimates of the CR for high-rate communication. 
In the EKF framework, we derive a state evolution model for the CR by incorporating wave-induced disturbance loads into the surge, sway, and yaw dynamics of the USV.
In addition, we further analyze the impacts of sea clutter and sea state on the measurement equation, which is constructed using observations from multiple scatterers arising from the extended-target characteristics of the USV.
Finally, we formulate a time allocation optimization problem to balance the time durations of the wide- and narrow-beam modes under different sea states, considering both robust USV state sensing and high-rate communication.
Simulation results demonstrate that the proposed scheme achieves consistently good tracking performance compared with conventional and neural-network-based methods.

\end{abstract}
	
	
	\section{INTRODUCTION}
	
W{\scshape ith} the rapid development of
maritime surveillance and smart ports,
maritime scenarios have placed high demands
on wireless communications for mobile
platforms such as uncrewed surface
vehicles (USVs), including low latency
and stable connectivity
\cite{10163065,9016362}.
However, complex marine environments, including tides, waves, and evaporation ducts, render maritime channels highly dynamic and nonlinear, posing significant challenges to the establishment of stable communication links \cite{10452297,11536108}. 
Conventional shore-based base stations suffer from limited coverage, whereas satellite communications usually entail high latency \cite{11614933}. 
This has motivated the deployment of uncrewed aerial vehicles (UAVs) as flexible aerial nodes to enhance the coverage, reliability, and overall performance of maritime communications~\cite{11455194,11219291,11586666}.
	
To enhance the transmission capability of UAV-assisted maritime communication systems, beamforming-enabled multi-antenna techniques have been widely adopted \cite{10246397}. 
By concentrating energy toward the desired direction, beamforming can effectively suppress interference and improve both the energy efficiency and spectral efficiency of the system \cite{8869849}. 
For example, the work in~\cite{10233419} investigates transmit beamforming in UAV-assisted maritime environmental monitoring networks and demonstrates that the adoption of beamforming can provide stronger system stability and significantly improve spectral efficiency.
However, such beam alignment methods typically rely on frequent beam training, where pilot signals are transmitted at the beginning of each time slot to perform beam sweeping or codebook search for determining the optimal beam direction \cite{11095988}.
As a result, frequent pilot transmission consumes a considerable amount of valuable time-frequency resources, thereby reducing the overall efficiency of data transmission \cite{11346817}.
	
With the development of beam alignment techniques, related studies have gradually shifted from the high-overhead paradigm of exhaustive codebook scanning to low-overhead frameworks centered on beam search and beam prediction \cite{10382696,10786908}. 
Among these methods, beam search aims to enhance beam acquisition efficiency by effectively reducing the search space and strategically optimizing beam training procedures to achieve reliable and accurate beam alignment~\cite{11072154}.
For example, the work in~\cite{11080296} investigates ideal beam patterns and proposes a low-complexity codebook design scheme based on second-order cone programming (SOCP) to mitigate beam misalignment during hierarchical beam training.
In contrast, beam prediction places greater emphasis on exploiting historical channel information to enable proactive and continuous beam alignment~\cite{10695819}.
Some studies directly employ deep learning for beam prediction, but such methods usually require large scale training data, suffer from limited generalization capability, and incur high hardware costs~\cite{10241295}.
Nevertheless, such methods still depend on training and beam scanning, resulting in non-negligible resource overhead \cite{7406682}.
	
	\begin{figure}[t]
		\centering
		\includegraphics[width=7cm]{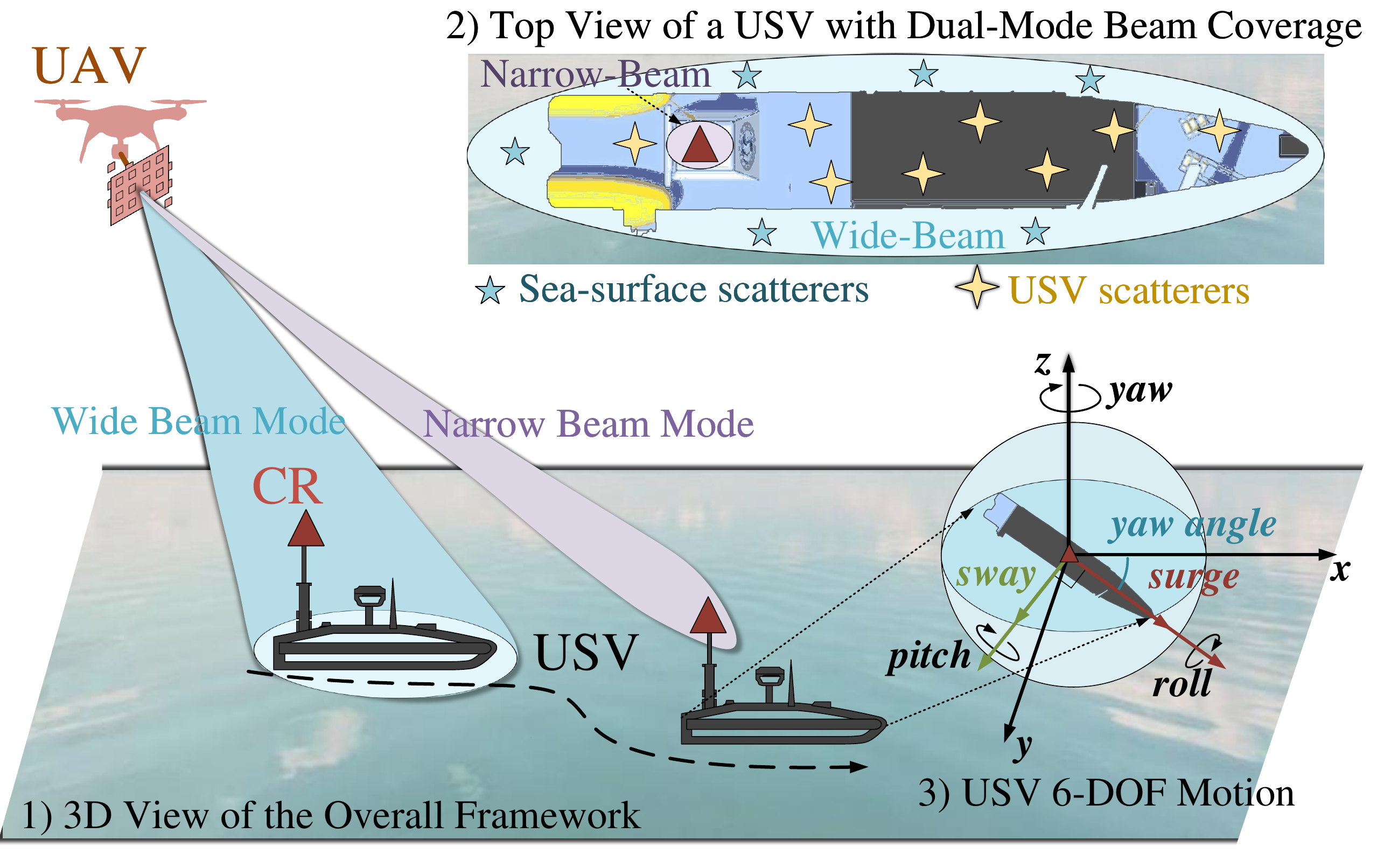}
		\caption{
			Sea-air beam tracking framework.
		}
		\label{fig1}
	\end{figure}

	Another line of work is based on sensing-assisted communication, where target state information is acquired during data transmission to support predictive beam tracking \cite{11534831,10705081}. 
	Building upon this concept, several studies have applied the integrated sensing and communication (ISAC) framework directly to beam alignment problems \cite{10279351,9246715}. 
	Sensing-assisted beam tracking enabled by ISAC offers three notable benefits, namely reduced pilot overhead, improved state estimation accuracy, and additional degrees of freedom for beam design~\cite{10561505}.
	In \cite{7888145}, a dedicated radar sensor is introduced in millimeter-wave vehicle-to-infrastructure (V2I) communications to assist beamforming, achieving more accurate beam alignment and reduced training overhead at the cost of additional hardware. 
	To avoid hardware modifications, the work in~\cite{9171304} proposed an ISAC-based predictive beam tracking scheme that estimates vehicle states from reflected communication echoes and employs an extended Kalman filter (EKF) to predict vehicle positions for beam alignment.
	Nevertheless, the scenario considered in~\cite{9171304} is limited to vehicles moving in a constant-velocity straight line.
	Furthermore, the work in~\cite{10543024} took into account more realistic vehicle motions in practical environments, where vehicles may perform overtaking, acceleration, and lane-changing maneuvers in addition to regular driving.
	In \cite{9947033}, a point-like vehicle target is modeled as an extended target with multiple scattering centers, and the position of the communication receiver (CR) is estimated and tracked within an ISAC framework.
Therefore, existing sensing-assisted beam tracking
studies mainly focus on terrestrial vehicular
scenarios, while their application to sea-air
communication networks faces several key challenges:
	\begin{itemize}
		\item 
		\textit{\textbf{USV motion with multiple degrees of freedom.}}  
		Under the combined effects of surge, sway, and yaw, the USV exhibits strongly coupled motion with multiple degrees of freedom, making conventional methods based on simplified kinematic models inadequate for accurate state prediction and CR tracking \cite{11402881}.
		\item
		\textit{\textbf{Motion disturbances caused by sea conditions.}}  
	Dynamic sea conditions introduce pronounced nonlinearities and uncertainties into USV motion, resulting in a substantial mismatch between simplified motion models (e.g., \cite{9947033}) and the actual time-varying USV dynamics, thereby degrading the accuracy, stability, and reliability of beam tracking.
		\item
		\textit{\textbf{Measurement noise caused by sea conditions.}}  
		Under dynamic sea conditions, radar echoes scattered from the sea surface give rise to sea clutter, which interferes with the echoes from distributed scatterers on the USV, thereby reducing the sensing signal-to-clutter-plus-noise ratio (SCNR) and degrading state estimation and beam tracking performance~\cite{8861449}.
	\end{itemize}
	These challenges motivate the design of a dedicated and robust beam tracking scheme for sea-air communication.

	\begin{figure}[t]
	\centering
	\includegraphics[width=8cm]{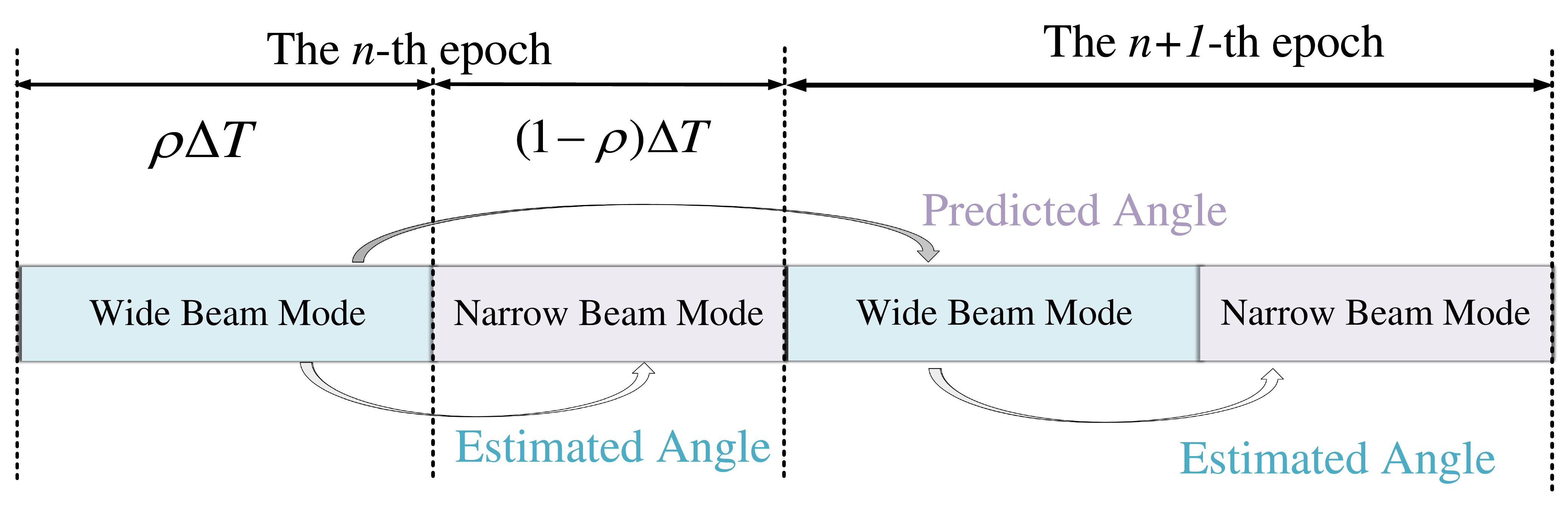}
	\caption{
		Dual-mode timing diagram
	}
	\label{fig2}
\end{figure}
\subsection{Contributions}

This paper proposes a sensing-assisted beam tracking framework for sea-air communication systems, in which the UAV communicates with the CR mounted on the USV while simultaneously sensing the USV state, as illustrated in Fig.~\ref{fig1}.
To characterize the physical reality of the USV, the USV is modeled as an extended scattering target with multiple distributed scatterers rather than a point target.
For robust beam tracking under dynamic sea
conditions while maintaining a high communication
rate, we divide each epoch into two stages
according to the time allocation factor,
namely the \emph{wide beam mode} and
the \emph{narrow beam mode}, as illustrated
in Fig.~\ref{fig2}.
The main contributions are summarized as follows in four aspects:
\begin{itemize}
	\item First, we propose a predictive beam tracking framework for sea-air communication networks, where the USV is explicitly modeled as an extended target. 
	A dual-mode scheme is developed to integrate wide-beam sensing and narrow-beam communication. 
	Specifically, the wide beam illuminates the entire extended USV for state estimation, while the narrow beam exploits the predicted CR state to achieve high-gain communication.
	
	\item 
	Second, we develop a sea-air tailored EKF model to track the six dimensional state of the CR on the USV. The state evolution equation explicitly accounts for the surge, sway, yaw, and sea induced disturbances of the USV. 
Then, the measurement equation for the extended target USV is formulated by incorporating the sea clutter induced measurement uncertainty into the corresponding covariance matrix.
	
	\item 
	Third, we formulate a sea condition based time allocation problem to achieve a trade-off between sensing accuracy and communication rate, where the posterior angular error covariance under sea clutter is used to characterize the impact of sensing time on beam alignment probability.
	
	\item 
	Finally, simulation results show that (i) the range dependent applicability boundary of the extended target model is clearly revealed, (ii) the effect of sea clutter intensity on wide beam sensing time allocation is revealed, and (iii) the proposed framework achieves reliable beam tracking and maintains more stable communication throughput under unseen sea conditions than learning-based benchmark schemes.

\end{itemize}

\textbf{\textit{Notations:}} Bold uppercase/lowercase letters denote matrices/vectors, respectively, while normal letters and calligraphic letters denote scalars and sets. 
$\|\cdot\|$ denotes the Euclidean norm, and $\mathbb{C}$ is the complex field. 
$(\cdot)^H$ denotes the Hermitian transpose, and $\mathbf{I}_M\in\mathbb{R}^{M\times M}$ is the identity matrix.
$\mathcal{N}(\mu, \sigma^2)$ is the normal distribution with mean $\mu$ and variance $\sigma^2$. 
$\mathcal{U}(0, 2\pi)$ denotes a uniform distribution in the interval $[0, 2\pi]$. $\otimes$ denotes the Kronecker product.

\section{System Model}	

In this paper, we consider a sea-air communication scenario, where a UAV is deployed at a fixed altitude $H$ and equipped with a uniform planar array (UPA) consisting of $N_x$ and $N_y$ antennas along the $x$- and $y$-axes, respectively, while a USV navigates on the sea surface, as illustrated in Fig.~\ref{fig1} \footnote{The UAV is assumed to hover during each beam-tracking interval to maintain stable beam transmission and avoid additional Doppler shifts caused by UAV motion \cite{11536108}.
	}.
The UAV maintains a continuous communication link with the CR on the USV, where the CR is equipped with a single antenna. To enable accurate tracking, the time interval $t\in[0,T]$ is divided into $N$ epochs, each with duration $\Delta T=T/N$, where $T$ is the maximum duration. 
At the $n$-th epoch, during the transmission phase, only a subset of the antenna elements is activated for beamforming. The corresponding transmit sub-array size is given by $N_{\mathrm{tx},n}=N_{x,n}N_{y,n}$, where the subscript $n$ indicates that the transmit sub-array configuration can be adaptively reconfigured at each epoch according to the beamwidth requirement.
During the signal reception phase, all antenna elements of the UPA are utilized for signal acquisition, yielding a receive array size of $N_{\mathrm{rx}} = N_x N_y$.
For clarity and ease of subsequent analysis, we define the CR state vector at the
$n$-th epoch as
$\mathbf{x}_n \triangleq
[\theta_n,\varphi_n,d_n,
v_{x,n},v_{y,n},v_{z,n}]^T$
for beam tracking.
Here, $\theta_n$, $\varphi_n$, and $d_n$
denote the azimuth, elevation, and distance
of the CR relative to the UAV, respectively,
while $v_{x,n}$, $v_{y,n}$, and $v_{z,n}$
denote its velocity components in the
inertial coordinate system.

\subsection{ Signal Model }

As shown in Fig.~\ref{fig2}, each epoch consists of two transmission modes: a wide-beam mode for both USV sensing and CR communication, followed by a narrow-beam mode for CR communication only.
The transmitted signal at the $n$-th epoch of the UAV can be expressed by
\begin{equation}
	\mathbf r_n(t)=\mathbf f_n x_n(t),
	\label{eq:unified_tx_signal}
\end{equation}
where $\mathbf f_n$ and $x_n(t)$ denote the beamforming vector and the transmit signal, respectively.
Specifically, the transmit signal $x_n(t)$ is given by
\begin{equation}
	x_n(t)=
	\begin{cases}
		s_n(t), & \text{wide-beam mode},\\
		c_n(t), & \text{narrow-beam mode},
	\end{cases}
\end{equation}
where $s_n(t)$ denotes the ISAC signal for both sensing and communication, and $c_n(t)$ denotes the communication signal.
The transmit beamforming vector $\mathbf f_n$ is given by
\begin{equation}
	\mathbf f_n=
	\begin{cases}
		\mathbf f_n^{\mathrm{wide}}, & \text{wide-beam mode},\\
		\mathbf f_n^{\mathrm{narr}}, & \text{narrow-beam mode},
	\end{cases}
	\label{eq:mode_dependent_tx}
\end{equation}
where $\mathbf f_n^{\mathrm{wide}}\in\mathbb C^{N_{x,n}N_{y,n}}$ and $\mathbf f_n^{\mathrm{narr}}\in\mathbb C^{N_xN_y}$ denote the beamforming vectors for the subarray and full UPA, respectively.
Next, we present the sensing signal model for the wide-beam mode and the communication signal model for both modes.
\subsubsection{Sensing Signal Model}

In the wide-beam mode, the UAV transmits the ISAC signal $s_n(t)$ to sense the extended USV and its surrounding environment. As shown in Fig.~\ref{fig1}, the echo signal received at the UAV consists of reflections from the $K$ USV scatterers and sea clutter, and is expressed as
\begin{equation}
	\notag
	\begin{aligned}
		&\mathbf y_n(t)
		=\kappa_n\sqrt p\sum_{k=1}^{K}
		\beta_{k,n}e^{j2\pi\mu_{k,n}t}\mathbf b(\theta_{k,n},\varphi_{k,n})
		\\&\times\mathbf a_{N_{\mathrm{tx},n}}^H(\theta_{k,n},\varphi_{k,n})\mathbf f_n^{\mathrm{wide}}s_n(t-\tau_{k,n})
		+\mathbf i_n^{\mathrm{cl}}(t)+\mathbf z_r(t),
	\end{aligned}
\end{equation}
where $\tau_{k,n}$, $\mu_{k,n}$, and $\beta_{k,n}$ denote the round-trip propagation delay, Doppler frequency shift, and effective complex channel coefficient of the $k$-th USV scatterer, respectively. $\kappa_n$ represents the sensing gain scaling factor, $p$ denotes the transmit power, $\mathbf i_n^{\mathrm{cl}}(t)$ denotes the aggregate sea-clutter echoes, and $\mathbf z_r(t)$ represents the receiver thermal noise.

The sensing signal model accounts for the effects of the maritime environment through two components: 1) the sensing channel coefficient $\beta_{k,n}$ of each USV scatterer and 2) the sea-clutter echoes $\mathbf i_n^{\mathrm{cl}}(t)$.
First, the round-trip sensing channel coefficient is modeled as
$
	\beta_{k,n}
	={\varepsilon_k}{(c\tau_{k,n})^{-2}}
	g^{\mathrm{sen}}_{k,n},
$
where $\varepsilon_k$ is the complex scattering amplitude whose squared magnitude is proportional to the radar cross section (RCS), $(c\tau_{k,n})^{-2}$ represents the round-trip path loss, and $g^{\mathrm{sen}}_{k,n}$ accounts for additional maritime propagation effects. 
Motivated by the maritime propagation characteristics discussed
in \cite{9460824}, the maritime propagation factor is approximated as
\begin{equation}
	g^{\mathrm{sen}}_{k,n}
	=10^{-X^{\mathrm{sen}}_{k,n}/20}h^{\mathrm{sen}}_{k,n},
	\label{eq:sensing_gain}
\end{equation}
where $X^{\mathrm{sen}}_{k,n}\sim\mathcal N(\mu_{\mathrm{sen}},\sigma_{\mathrm{sen}}^2)$ represents the slowly varying attenuation in dB. 
The Rician fading coefficient $h^{\mathrm{sen}}_{k,n}$
for the $k$-th scatterer is given by
\begin{equation}
	h^{\mathrm{sen}}_{k,n}
	=\sqrt{\frac{K^{\mathrm{sen}}_{k,n}}{K^{\mathrm{sen}}_{k,n}+1}}
	+\sqrt{\frac{1}{K^{\mathrm{sen}}_{k,n}+1}}
	\tilde h^{\mathrm{sen}}_{k,n},
	\label{eq:sensing_rician}
\end{equation}
where $\tilde h^{\mathrm{sen}}_{k,n}\sim\mathcal{CN}(0,1)$ and $K^{\mathrm{sen}}_{k,n}$ is the Rician factor.
Second, the impact of sea clutter on sensing accuracy is further analyzed in Sec.III-\ref{B-2}.

In addition, $\mathbf{a}_{N_{\mathrm{tx},n}}(\theta,\varphi)$ and $\mathbf{b}(\theta,\varphi)$
denote the transmit and receive steering vectors of the UAV's UPA, 
respectively. 
$\mathbf{a}_{N_{\mathrm{tx},n}}(\theta,\varphi)$
denotes the steering vector of the transmit
subarray with $N_{\mathrm{tx},n}$ active antennas
at the $n$-th epoch.
For compact notation, define the normalized steering vector of a ULA with $M$ elements as 
\begin{equation} 
	\mathbf{v}_M(x) 
	= 
	\frac{1}{\sqrt{M}} 
	[ 
	1,e^{j\pi x},\ldots,e^{j\pi(M-1)x} 
	]^T . 
\end{equation} 
Then, the transmit steering vector is given by 
\begin{equation}  
	\mathbf{a}_{N_{\mathrm{tx},n}}(\theta,\varphi)  
	=  
	\mathbf{v}_{N_{x,n}}(\sin\varphi\cos\theta)  
	\otimes  
	\mathbf{v}_{N_{y,n}}(\sin\varphi\sin\theta). 
	\notag
\end{equation}  
The receive steering vector is given by
\begin{equation} 
	\mathbf{b}(\theta,\varphi) 
	= 
	\mathbf{v}_{N_{x}}(\sin\varphi\cos\theta) 
	\otimes 
	\mathbf{v}_{N_{y}}(\sin\varphi\sin\theta). 
\end{equation} 

Then, let $\widehat{\theta}_{n|n-1}$ and $\widehat{\varphi}_{n|n-1}$ denote the predicted azimuth and elevation angles of the CR, respectively, which are obtained from the state estimate at the $(n-1)$-th epoch using the state evolution equations developed in Sec.III-\ref{secIIIA}. 
Accordingly, the transmit beamforming vector is constructed by exploiting the predicted angular direction of the CR as
$\mathbf{f}^{\mathrm{wide}}_n=\mathbf{a}_{N_{\mathrm{tx},n}}(\widehat{\theta}_{n|n-1},\widehat{\varphi}_{n|n-1})$.

\subsubsection{Communication Signal Model}

The UAV keeps communicating with the single-antenna CR in both beam modes. Based on \eqref{eq:unified_tx_signal}, the received signal at the CR is given by
\begin{equation}
	y_n(t)
	=\alpha_n\sqrt{N_{\mathrm{tx},n}p}\,
	{\mathbf a}_{N_{\mathrm{tx},n}}^H(\theta_n,\varphi_n)
	\mathbf f_n x_n(t)+z_n^C(t),
	\label{eq:unified_comm_rx}
\end{equation}
where $\alpha_n$ denotes the UAV-CR channel coefficient, and $z_n^C(t)\sim\mathcal{CN}(0,\sigma_c^2)$ is the receiver noise. 
Similarly, the communication channel which accounts for maritime propagation effects can be modeled as
\begin{equation}
	\alpha_n
	=\alpha_{\mathrm{ref}}d_n^{-1}
	10^{-A_n^{\mathrm{com}}/20}h_n^{\mathrm{com}}
	e^{-j\frac{2\pi f_c}{c}d_n},
	\label{eq:maritime_comm_channel}
\end{equation}
where $\alpha_{\mathrm{ref}}$ is the reference channel coefficient, $d_n^{-1}$ represents the distance-dependent attenuation, and $A_n^{\mathrm{com}}$ denotes the additional maritime attenuation in dB. The Rician fading coefficient $h_n^{\mathrm{com}}$ is given by
\begin{equation}
	h_n^{\mathrm{com}}
	=\sqrt{\frac{K^{\mathrm{com}}}{K^{\mathrm{com}}+1}}
	+\sqrt{\frac{1}{K^{\mathrm{com}}+1}}
	\tilde h_n^{\mathrm{com}},
	\label{eq:communication_rician}
\end{equation}
where $\tilde h_n^{\mathrm{com}}\sim\mathcal{CN}(0,1)$
and $K^{\mathrm{com}}$ denotes the Rician fading factor
of the communication channel.

\textit{1) Wide Beam Mode:}
In the wide-beam mode, the UAV employs a transmit subarray for simultaneous USV sensing and CR communication, namely, \(N_{\mathrm{tx},n}=N_{x,n}N_{y,n}\) active antennas.
The corresponding received
SNR at the CR is given by
\begin{equation}
	\mathrm{SNR}_n^{\mathrm{wide}}
	=\frac{pN_{\mathrm{tx},n}|\alpha_n|^2
		|\mathbf a_{N_{\mathrm{tx},n}}^H(\theta_n,\varphi_n)
		\mathbf f_n^{\mathrm{wide}}|^2}{\sigma_c^2}.
	\label{Wide}
\end{equation}
The corresponding achievable communication rate is given by $R_n^{\mathrm{wide}}=\log_2(1+\mathrm{SNR}_n^{\mathrm{wide}})$.

\textit{2) Narrow Beam Mode:}
In the narrow-beam mode, the UAV uses the full UPA
for CR communication, with the beam directed toward
the CR based on the updated angle estimates,
namely, $N_{\mathrm{tx},n}=N_xN_y$.
The corresponding received SNR at the CR is then given by
\begin{equation}
	\mathrm{SNR}_n^{\mathrm{narr}}
	=\frac{pN_{\mathrm{tx},n}|\alpha_n|^2
		|{\mathbf a}_{N_{\mathrm{tx},n}}^H(\theta_n,\varphi_n)
		\mathbf f_n^{\mathrm{narr}}|^2}{\sigma_c^2},
	\label{narr}
\end{equation}
where $\mathbf f_n^{\mathrm{narr}}
={\mathbf a}_{N_{\mathrm{tx},n}}
(\widehat{\theta}_n,\widehat{\varphi}_n)$,
and ($\widehat{\theta}_n$,
$\widehat{\varphi}_n$) denote the estimated
CR azimuth and elevation angles, respectively.
These estimates are obtained from the state
update in Step 6 of Algorithm~\ref{alg:ekf}.
The corresponding achievable rate is
$R_n^{\mathrm{narr}}=\log_2(1+\mathrm{SNR}_n^{\mathrm{narr}})$.
Since $N_xN_y\geq N_{x,n}N_{y,n}$, the narrow-beam mode is expected to achieve a higher overall communication rate than the wide-beam mode when the beam is accurately directed toward the USV.
Therefore, our objective is to maximize the available time allocated to narrow-beam communication while ensuring reliable and accurate tracking through wide-beam sensing.

\subsection{CR Motion Model}

Since the CR is mounted on the USV,
we consider four degrees of freedom (4-DoF)\footnote{Following the common practice in marine vessel modeling, we neglect roll and pitch motions under mild-to-moderate sea conditions, where their effects on CR motion and beam alignment are assumed to be relatively small~\cite{7098387}.},
with surge, sway, and yaw described by a
classical 3-DoF dynamic model~\cite{7098387}
and vertical motion modeled separately
in Sec.~III-\ref{secIIIA}.
Therefore,
to establish the 3-DoF
dynamic model of the USV, we first define
its body-fixed velocity state at the
$n$-th epoch as
$\boldsymbol{\nu}_n=[u_n,w_n,r_n]^T\in\mathbb{R}^{3\times 1}$,
where $u_n$, $w_n$, and $r_n$ denote the surge velocity, sway velocity, and yaw rate, respectively, as shown in Fig.~\ref{fig1}.\footnote{Since the CR is rigidly mounted on the USV, it shares the same body-fixed translational and rotational motions as the USV throughout the considered operating period. Therefore, $u_n$, $w_n$, and $r_n$ can be used to characterize the corresponding motion states of the CR.
}
Therefore, the 3-DoF dynamics are given by~\cite{7098387}\footnote{  Here, $\dot{\boldsymbol{\nu}}_n$ denotes the continuous-time derivative of the body-fixed velocity vector $\boldsymbol{\nu}$, evaluated at the $n$-th sampling instant.
 }
\begin{equation}
	\mathbf{M}\dot{\boldsymbol{\nu}}_n
	+\mathbf{C}\boldsymbol{\nu}_n
	+\mathbf{D}\boldsymbol{\nu}_n
	=\boldsymbol{\tau}^{\mathrm{con}}_n
	+\boldsymbol{\tau}^{\mathrm{sea}}_n,
	\label{3DofUSV}
\end{equation}
where $\mathbf{M}$, $\mathbf{C}$, and $\mathbf{D}$ denote the mass-inertia, Coriolis and centripetal, and hydrodynamic damping matrices, respectively. 
These matrices are determined by the USV's
physical properties and treated as known
system quantities~\cite{10033669}, with
their detailed expressions provided in
Appendix~\ref{app:usv_matrices}.
The control input and sea-induced disturbance load vectors are defined as
$\boldsymbol{\tau}^{\mathrm{con}}_n=
[\tau^{\mathrm{con}}_{1,n},\tau^{\mathrm{con}}_{2,n},\tau^{\mathrm{con}}_{3,n}]^T$
and $\boldsymbol{\tau}^{\mathrm{sea}}_n= [\tau^{\mathrm{sea}}_{1,n},\tau^{\mathrm{sea}}_{2,n},\tau^{\mathrm{sea}}_{3,n}]^T$, respectively. The three components of each vector represent the surge force, sway force, and yaw moment, respectively.

To characterize the effects of sea waves, the sea-induced disturbance loads are modeled as~\cite{wicaksono2018wave}
\begin{equation}
	\boldsymbol{\tau}^{\mathrm{sea}}_n
	=\rho_w g a_n^2
	\boldsymbol{\mathcal{G}}
	\left(
	k_{0,n},\omega_{e,n},\chi_n,U_n,
	\boldsymbol{\Theta}_{\mathrm{usv}}
	\right),
	\label{hailang}
\end{equation}
where $\rho_w$, $g$, and $a_n$ denote the water density, gravitational acceleration, and wave amplitude, respectively. The function $\boldsymbol{\mathcal{G}}(\cdot)$ represents the hydrodynamic load model, which depends on the wave number $k_{0,n}$, encounter frequency $\omega_{e,n}$, relative wave direction $\chi_n$, USV speed $U_n$, and fixed USV parameters $\boldsymbol{\Theta}_{\mathrm{usv}}$.\footnote{
The USV motion information is obtained from its onboard sensors, while the sea-wave parameters are obtained from meteorological data or onboard navigation radar. These parameters describe sea conditions averaged over a given time window and are used as prior information for the sea-induced disturbance model. However, such statistical information alone cannot accurately capture the instantaneous USV motion required for reliable real-time beam tracking, which motivates us to develop an EKF-based beam tracking method.
} 
The detailed expressions of the wave-induced load function $\boldsymbol{\mathcal G}(\cdot)$ are provided in Appendix~\ref{app:wave_load}.
With the known USV control inputs
$\boldsymbol{\tau}^{\mathrm{con}}_n$
and estimated sea-induced disturbance loads
$\boldsymbol{\tau}^{\mathrm{sea}}_n$,
the USV dynamic model in \eqref{3DofUSV}
describes the coupled evolution of the
body-fixed velocity $\boldsymbol{\nu}_n$
under dynamic sea conditions, thereby
providing a physical basis for deriving
the CR velocity evolution equations
in Sec.~III-\ref{secIIIA}.

\section{ISAC-Based Beam Tracking Scheme}
\label{secIII}

In this section, we first derive the state evolution equation of the CR. Second, we formulate the measurement equation under the impact of sea clutter and present the overall EKF framework. Then, we establish a time allocation optimization problem and derive a suboptimal solution for the time allocation factor. 

\subsection{State Evolution Model}	
\label{secIIIA}


Our objective is to predict the CR state $\mathbf{x}_n$ at the \(n\)-th epoch based on the information available at the \((n-1)\)-th epoch. 
To predict the CR state, we transform the CR body-fixed velocities $\boldsymbol{\nu}_n$ into inertial-frame components $v_{x,n}$, $v_{y,n}$, and $v_{z,n}$, and use this velocity evolution to derive the angular and range state equations. The detailed derivation follows.

\subsubsection{CR Velocity Evolution Model}
For a sufficiently small sampling interval,
the yaw rate $r_n$ can be regarded as
locally constant within each epoch, i.e.,
$r_n \approx r_{n-1}$, which gives the
heading-angle increment
$\Delta\psi_n \approx r_{n-1}\Delta T$,
where $\psi_n$ denotes the heading angle.\footnote{
This local discretization approximation
assumes an approximately constant yaw
rate within each epoch, provided that
the sampling interval is much shorter
than the characteristic time scale of
sea-induced yaw-rate variations, such as
the dominant wave period.
It does not imply negligible yaw
acceleration over the entire motion process.
}
Therefore, by expanding \eqref{3DofUSV}, the motion equations can be written as~\cite{10033669}
\begin{equation}
	\left\{
	\begin{aligned}
		& m_{11}\dot{u}_n + c_{13}r_n + d_{11}u_n
		= \tau^{\mathrm{con}}_{1,n} + {\tau}^{\mathrm{sea}}_{1,n}, \\
		& m_{22}\dot{w}_n + c_{23}r_n + d_{22}w_n + d_{23}r_n
		= \tau^{\mathrm{con}}_{2,n} + {\tau}^{\mathrm{sea}}_{2,n}.
	\end{aligned}
	\right.
	\label{reduced_surge_sway_dynamics}
\end{equation}
Since the velocity measurements are defined in the inertial Cartesian frame, the body-fixed velocities are transformed into inertial-frame horizontal components as
\begin{equation}
	\begin{bmatrix}
		v_{x,n}\\
		v_{y,n}
	\end{bmatrix}
	=
	\begin{bmatrix}
		\cos\psi_n & -\sin\psi_n\\
		\sin\psi_n & \cos\psi_n
	\end{bmatrix}
	\begin{bmatrix}
		u_n\\
		w_n
	\end{bmatrix}.
	\label{velocity_transform}
\end{equation}

Taking the time derivative of \eqref{velocity_transform} and applying the product rule to the rotation matrix, the inertial-frame acceleration transformation is obtained as
\begin{equation}
	\begin{bmatrix}
		\dot v_{x,n}\\
		\dot v_{y,n}
	\end{bmatrix}
	=
	\begin{bmatrix}
		\cos\psi_n & -\sin\psi_n\\
		\sin\psi_n & \cos\psi_n
	\end{bmatrix}
	\begin{bmatrix}
		\dot u_n\\
		\dot w_n
	\end{bmatrix}
	+
	r_n
	\begin{bmatrix}
		-v_{y,n}\\
		v_{x,n}
	\end{bmatrix}.
	\label{inertial_acc_transform_app}
\end{equation}

Substituting \eqref{reduced_surge_sway_dynamics} and \eqref{velocity_transform} into \eqref{inertial_acc_transform_app} yields the explicit acceleration expressions in \eqref{shizi_matrix}.
In \eqref{shizi_matrix}, for notational compactness, the repeated trigonometric and parameter terms are defined as \(\varsigma_n\triangleq\cos\psi_n\), \(\zeta_n\triangleq\sin\psi_n\), \(\lambda_1\triangleq d_{11}/m_{11}\), \(\lambda_2\triangleq d_{22}/m_{22}\), \(\lambda_3\triangleq c_{13}/m_{11}\), and \(\lambda_4\triangleq (c_{23}+d_{23})/m_{22}\).
The derived horizontal accelerations are then used to obtain the discrete-time velocity evolution model via first-order approximation.

\begin{figure*}[!b]
	\hrulefill
	\begin{equation}
		\label{shizi_matrix}
		\begin{bmatrix}
			\dot v_{x,n}\\
			\dot v_{y,n}
		\end{bmatrix}
		=
		\begin{bmatrix}
			-\lambda_1 \varsigma_n^2-\lambda_2 \zeta_n^2
			&
			(\lambda_2-\lambda_1)\zeta_n\varsigma_n-r_n
			\\
			(\lambda_2-\lambda_1)\zeta_n\varsigma_n+r_n
			&
			-\lambda_1 \zeta_n^2-\lambda_2 \varsigma_n^2
		\end{bmatrix}
		\begin{bmatrix}
			v_{x,n}\\
			v_{y,n}
		\end{bmatrix}
		+
		\begin{bmatrix}
			-\lambda_3 \varsigma_n+\lambda_4 \zeta_n\\
			-\lambda_3 \zeta_n-\lambda_4 \varsigma_n
		\end{bmatrix}
		r_n
		+
		\begin{bmatrix}
			\frac{\varsigma_n}{m_{11}} & -\frac{\zeta_n}{m_{22}}\\
			\frac{\zeta_n}{m_{11}} & \frac{\varsigma_n}{m_{22}}
		\end{bmatrix}
		\begin{bmatrix}
			\tau^{\mathrm{con}}_{1,n}+{\tau}^{\mathrm{sea}}_{1,n}\\
			\tau^{\mathrm{con}}_{2,n}+{\tau}^{\mathrm{sea}}_{2,n}
		\end{bmatrix}.
	\end{equation}
\end{figure*}

In addition, in practical maritime environments, the vertical velocity typically varies much more slowly than the horizontal velocity components over a short sampling interval. Hence, it is modeled as a first-order random-walk process, i.e., $v_{z,n}=v_{z,n-1}+\omega_z$, where $\omega_z\sim\mathcal{N}(0,\sigma_z^2)$ accounts for the residual uncertainty arising from the neglected vertical dynamics~\cite{5603589}. Finally, the complete CR velocity evolution model can be summarized as
\begin{equation}\label{Fin11}
	\begin{cases}
		v_{x,n} = v_{x,n-1} + \dot v_{x,n-1} \Delta T + \omega_{x}, \\
		v_{y,n} = v_{y,n-1} + \dot v_{y,n-1} \Delta T + \omega_{y}, \\
		v_{z,n} = v_{z,n-1} + \omega_{z},
	\end{cases}
\end{equation}
where \(\omega_{x} \sim \mathcal{N}(0, \sigma_x^2)\), \(\omega_{y} \sim \mathcal{N}(0, \sigma_y^2)\), and \(\omega_{z} \sim \mathcal{N}(0, \sigma_z^2)\) represent the corresponding process noise terms for the velocities along the inertial \(x\)-, \(y\)-, and \(z\)-axes.
Based on the obtained velocity evolution model, we next derive the state evolution equations of the azimuth angle, elevation angle, and distance by using the UAV-CR geometric relationship.

\subsubsection{CR Angular Evolution Model}
\begin{figure*} [t]
	\centering
	\subfloat[\normalfont \label{3d1}3D view]{
		\includegraphics[width=5.5cm]{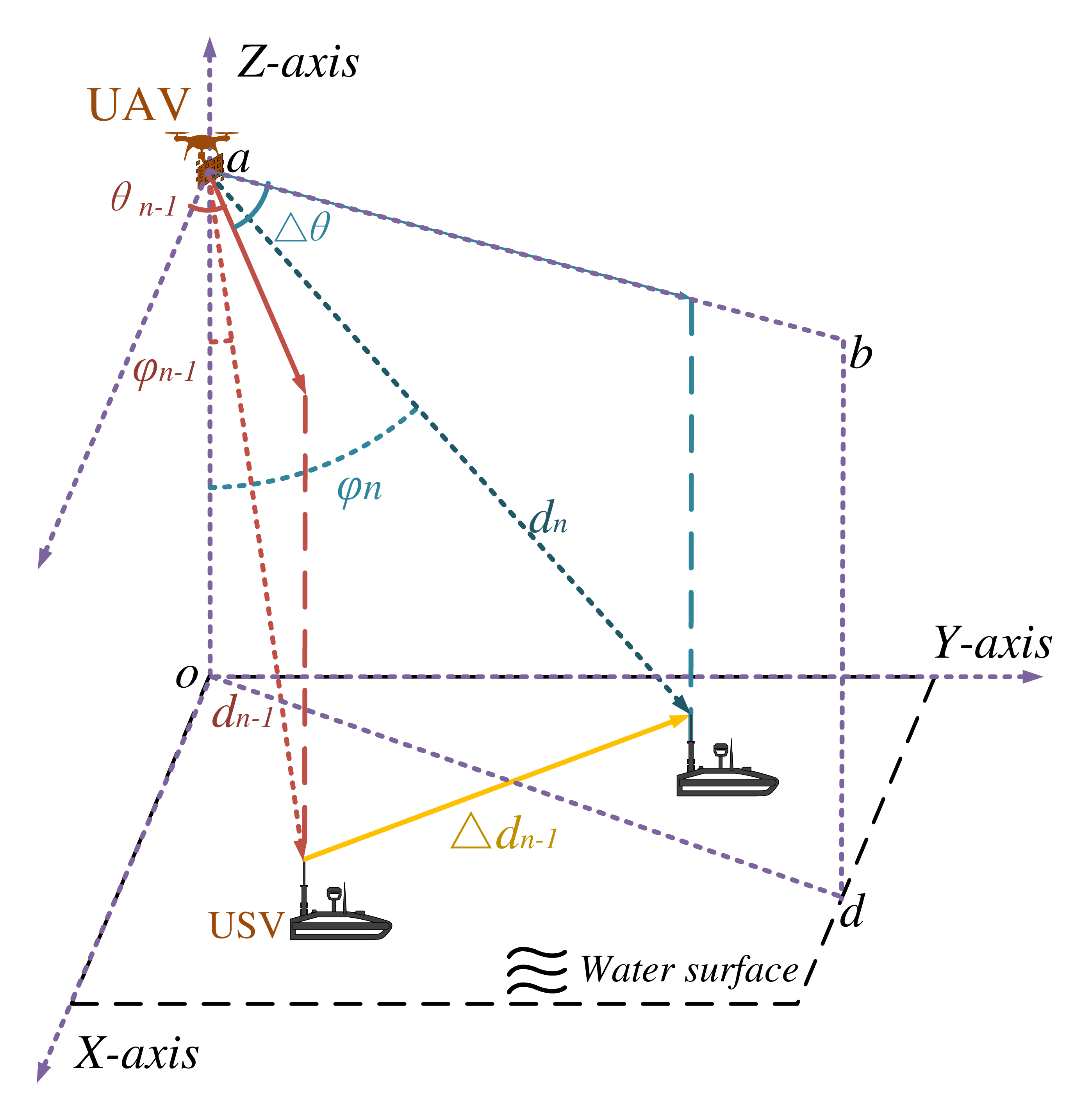}}
	\subfloat[\normalfont\label{3d2} XoY plane]{
		\includegraphics[width=5.5cm]{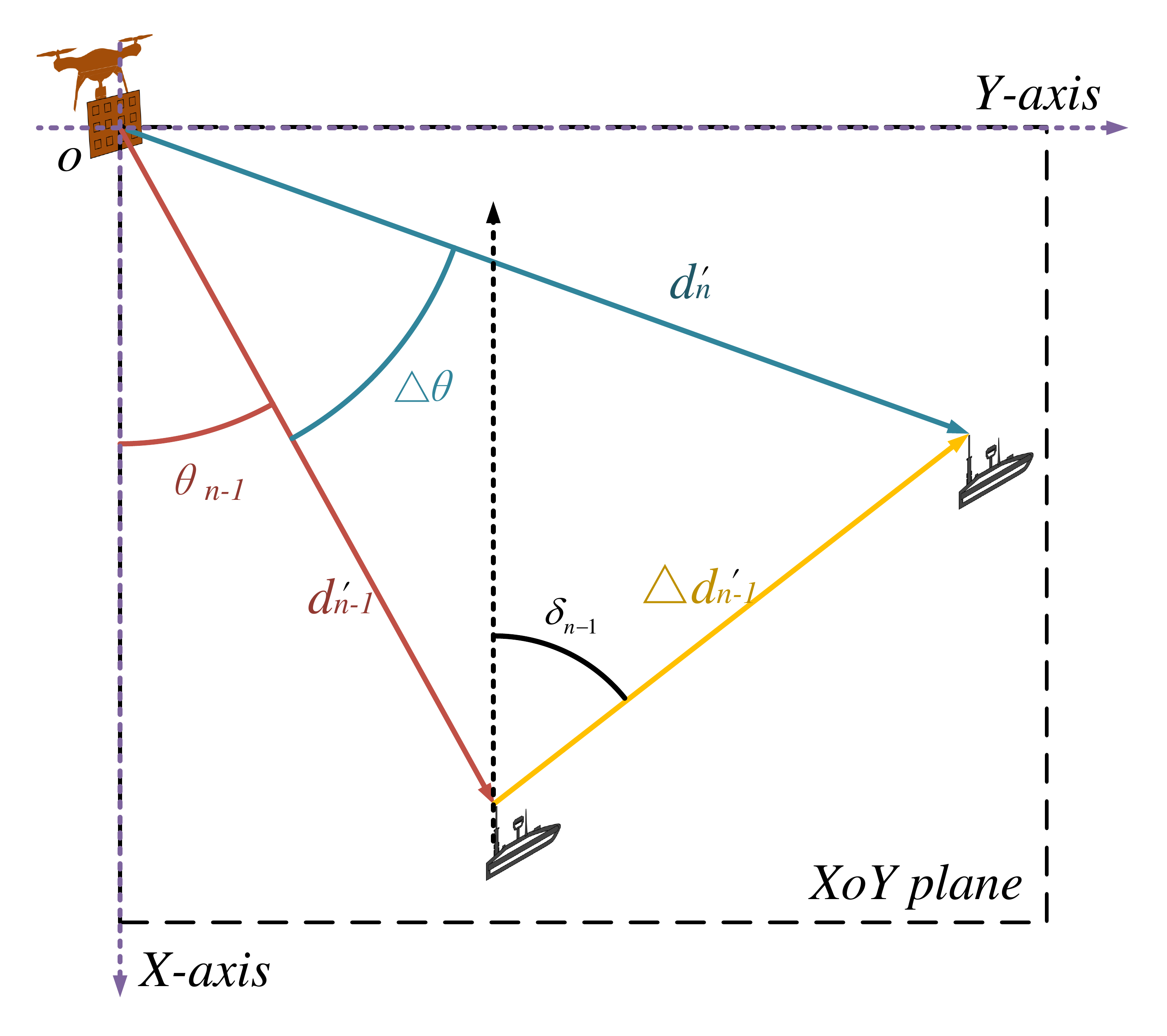}}
	\subfloat[\normalfont\label{3d3} $Oabd$ plane]{
		\includegraphics[width=5.5cm]{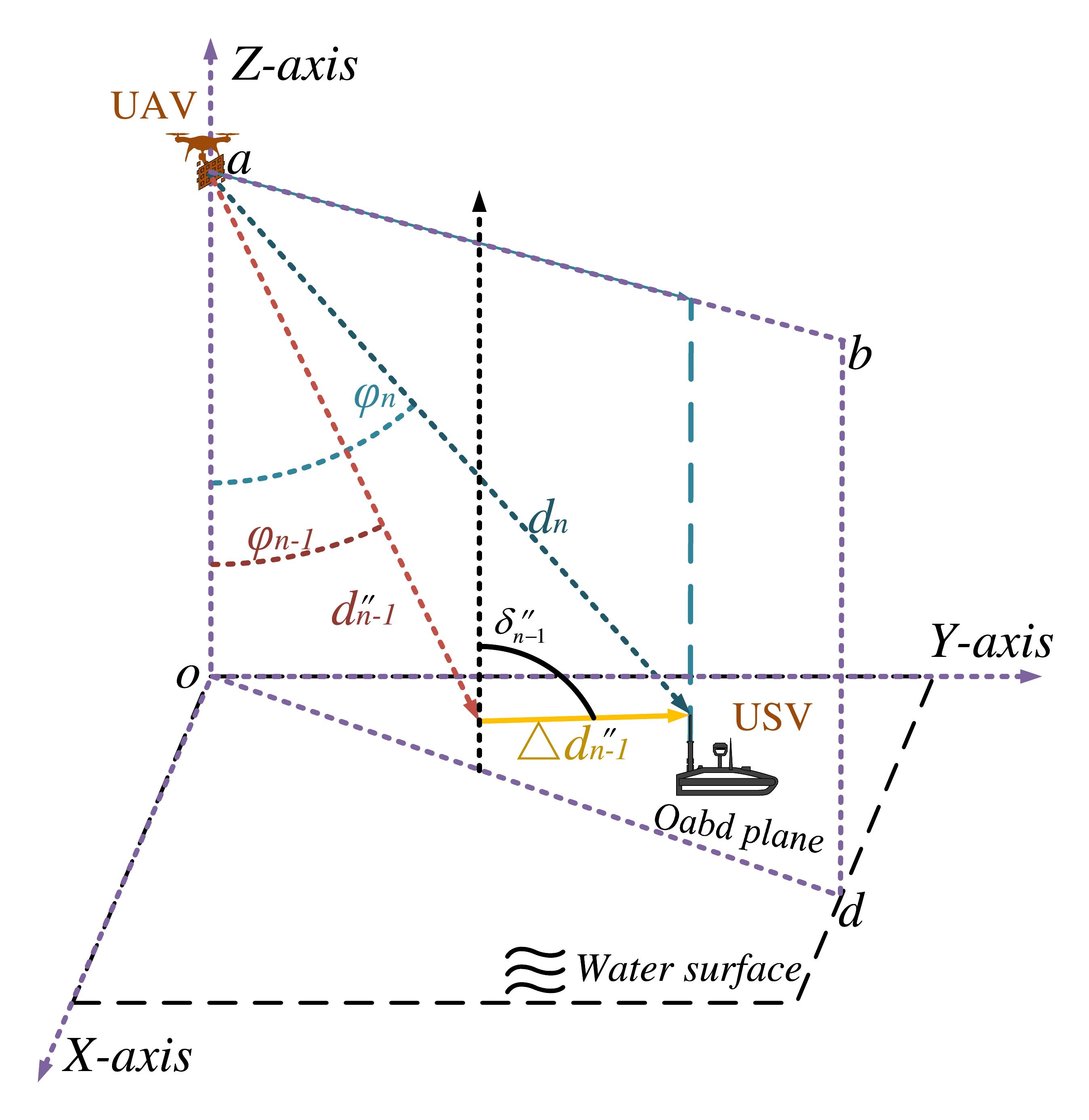}}
	\caption{The positional relationship between the CR on the USV and the UAV.}
	\label{We} 
\end{figure*}
\textit{1) Azimuth Evolution via Horizontal Projection:}
Since the direct analysis of the 3D geometry in Fig.~\ref{3d1} is complex, the geometric relation among $d_n$, $d_{n-1}$, and $\Delta d_{n-1}$ is projected onto the horizontal plane, as shown in Fig.~\ref{3d2}, to characterize the azimuth angle $\theta_n$. 
Here, $\Delta d_{n-1}$ denotes the displacement between the endpoints of $d_{n-1}$ and $d_n$. 
Using the inertial-frame accelerations $\dot{v}_{x,n-1}$ and $\dot{v}_{y,n-1}$ derived in the previous subsection, and accounting for acceleration effects, the components of $\Delta d_{n-1}$ along the $x$-, $y$-, and $z$-axes are given by
\begin{equation}
	\begin{aligned}
		&\Delta d_{x,n-1} = v_{x,n-1} \Delta T + \frac{1}{2} \dot v_{x,n-1} \Delta T^2, \\
		&\Delta d_{y,n-1} = v_{y,n-1} \Delta T + \frac{1}{2} \dot v_{y,n-1} \Delta T^2, \\
		&\Delta d_{z,n-1} = v_{z,n-1} \Delta T.
	\end{aligned}
\end{equation}
The total displacement is the Euclidean norm of its three components
$
\Delta d_{n-1}
= \sqrt{\smash[b]{\Delta d_{x,n-1}^{2}+\Delta d_{y,n-1}^{2}+\Delta d_{z,n-1}^{2}}}.
$

Now, to further analyze the azimuthal relationship, we focus on the horizontal projections of \( d_n \) and \( d_{n-1} \).
We denote the projections of $d_n$ and $d_{n-1}$ on the horizontal plane as $d_n'$ and $d_{n-1}'$, respectively, where $d_{n-1}'=d_{n-1}\sin\varphi_{n-1}$ and $d_n'=d_n\sin\varphi_n$.
Therefore, the relationship between the horizontal angle and the distance on the horizontal plane can be expressed by the cosine law and the sine law as
\begin{align}
	&	{d_n'}^2={d_{n-1}'}^2+\Delta {d'_{n-1}}^2-2{d_{n-1}'}\Delta {d'_{n-1}}\cos(\theta_{n-1} + \delta_{n-1}), \notag\\
	& \quad   \quad\quad\quad
	{\Delta {d'_{n-1}}}/{\sin\Delta\theta}={{d_n'}}/{\sin(\theta_{n-1} + \delta_{n-1})},
	\notag
\end{align}
where $\Delta\theta = \theta_n - \theta_{n-1}$ and $\delta_{n-1} = \arctan\left(\frac{\Delta d_{y,n-1}}{\Delta d_{x,n-1}}\right)$.
In addition, $\Delta d_{n-1}' = \sqrt{\smash[b]{\Delta d_{x,n-1}^2 +\Delta d_{y,n-1}^2}}$.
Then, following a geometric analysis similar to that in \cite{9171304}, the state evolution of \(d_n'\) and \(\theta_n\) can be obtained as
\begin{equation}
	\begin{cases}
		d_n'= d_{n-1}'-\Delta d_{n-1}'\cos(\theta_{n-1}+\delta_{n-1}), \\[0.5ex]
		\theta_n=\theta_{n-1}+{\Delta d_{n-1}'\sin(\theta_{n-1}+\delta_{n-1})}/{d_{n-1}'},
	\end{cases}
	\label{Fs4}
\end{equation}
where the intermediate derivation steps are omitted for brevity.
It is noted that the evolution of $\theta_n$ in \eqref{Fs4} is determined only by the known quantities at the $(n-1)$-th epoch.
Next, we derive the state evolution equations of the remaining two variables, i.e., the elevation angle $\varphi_n$ and the distance $d_n$.

\textit{2) Elevation and Range Evolution via Planar Mapping:}
Since $\varphi_{n-1}$, $d_{n-1}$, $\varphi_n$, and $d_n$ lie on two different planes, the plane containing $d_{n-1}$ is rotated about the $z$-axis onto the $Oabd$ plane. 
This preserves $\varphi_{n-1}$ and $d_{n-1}$, yielding the planar configuration shown in Fig.~\ref{3d3}. 
The elevation-angle relation on the $Oabd$ plane is given by
\begin{align}
	&{d^2_n}={d^2_{n-1}}+\Delta {d_{n-1}''}^2-2{d_{n-1}}\Delta {d_{n-1}''}\cos(\varphi_{n-1}+{\delta}_{n-1}''), \notag\\& \quad \quad \quad \quad 
	{\Delta {d_{n-1}}''}/{\sin\Delta\varphi}={{d_n}}/{\sin(\varphi_{n-1}+{\delta}_{n-1}'')}.
	\notag
\end{align}
Through a set-theoretic analysis, we obtain
\begin{equation}
	\begin{cases}
		d_n=d_{n-1}-\Delta d_{n-1}^{\prime\prime}\cos(\varphi_{n-1}+{\delta}_{n-1}''), \\
		\varphi_n=\varphi_{n-1}+{\Delta d_{n-1}^{\prime\prime}\sin(\varphi_{n-1}+{\delta}_{n-1}'')}{d^{-1}_{n-1}}. & 
	\end{cases}
	\label{Cs2}
\end{equation}
The derivation of \eqref{Cs2} follows a procedure similar to that in \eqref{Fs4} and is therefore omitted for brevity.
However, in \eqref{Cs2}, the expressions of $\Delta d_{n-1}^{\prime\prime}$ and ${\delta}_{n-1}''$ are still unknown at this stage, which motivates the following step.

First, before rotation, let $\mathbf d_{n-1}$ and $\mathbf d_n$ denote the displacement vectors from the UAV to the CR at epochs $n-1$ and $n$, respectively. They can be expressed as
\begin{equation}
	\mathbf d_i
	= d_i
	[
	\sin\varphi_i\cos\theta_i,
	\sin\varphi_i\sin\theta_i,
	-\cos\varphi_i
	]^T,
	\label{eq:d_vec}
\end{equation}
where $i\in\{n-1,n\}$.
Let $\gamma_n$ be the angle between $\mathbf d_{n-1}$ and $\mathbf d_n$. By the inner-product identity,
\begin{align}
	&\cos\gamma_n
	={\mathbf d_{n-1}^T\mathbf d_n}/({d_{n-1}d_n}) \notag\\
	&=\cos\varphi_{n-1}\cos\varphi_n
	+\sin\varphi_{n-1}\sin\varphi_n\cos(\Delta\theta),
	\label{eq:cos_gamma}
\end{align}
where $\Delta\theta = \theta_n-\theta_{n-1}$. 
Hence, the displacement between two consecutive CR positions is given by
\begin{align}
	\Delta d_{n-1}^{\,2}
	&=\|\mathbf d_n-\mathbf d_{n-1}\|^2 \notag\\
	&=d_n^2+d_{n-1}^2-2d_nd_{n-1}\cos\gamma_n .
	\label{eq:R2}
\end{align}

Second, after rotating the plane of $\mathbf d_{n-1}$ about the $z$-axis to match the azimuth of $\mathbf d_n$, denote the rotated vector by $\mathbf d_{n-1}''$. Then $\mathbf d_{n-1}''$ and $\mathbf d_n$ share the same azimuth, and their angular difference reduces to the elevation difference only, i.e.,
\begin{align}
	\cos\gamma_n''
	&=\cos(\Delta\varphi) \notag\\
	&=\cos\varphi_{n-1}\cos\varphi_n+\sin\varphi_{n-1}\sin\varphi_n ,
	\label{eq:cos_gamma_pp}
\end{align}
where $\Delta\varphi \triangleq \varphi_n-\varphi_{n-1}$. Therefore,
\begin{align}
	\Delta {d_{n-1}''}^{\,2}
	&=\|\mathbf d_{n}-\mathbf d_{n-1}''\|^2 \notag\\
	&=d_n^2+d_{n-1}^2-2d_nd_{n-1}\cos\gamma_n'' .
	\label{eq:R1}
\end{align}
Subtracting \eqref{eq:R2} from \eqref{eq:R1} and using
$1-\cos(\Delta\theta)=2\sin^2(\Delta\theta/2)$ yields
\begin{equation}
	\Delta {d_{n-1}''}^2
	={\Delta d_{n-1}^{\,2}
		-4d_{n-1}d_n\sin\varphi_{n-1}\sin\varphi_n\sin^2({\Delta\theta}/{2}}) . \label{xsw123}
\end{equation}

Next, using the horizontal projection relation $d_n'=d_n\sin\varphi_n$ derived in \eqref{Fs4}, we obtain
\begin{equation}
	d_n\sin\varphi_n= d_{n-1}\sin\varphi_{n-1}-\Delta d_{n-1}'\cos(\theta_{n-1}+\delta_{n-1}).
	\label{dnsinzxc}
\end{equation}

Based on the expression in \eqref{dnsinzxc} and by substituting it into \eqref{xsw123}, we can derive the compact representation of \( \Delta d_{n-1}'' \) as
\begin{equation} 		\label{qwerrewq}
	\Delta d_{n-1}'' = \sqrt{\smash[b]{\Delta d_{n-1}^2 - \tilde d^2_{n-1}}},
\end{equation}
where the definition of \( \tilde d^2_{n-1} \) is given in \eqref{eryuan1}.
Further, $\delta_{n-1}''$ denote the angle between the rotated $\Delta d_{n-1}''$ and the $z$-axis as 
\begin{equation}
	\delta_{n-1}''
	=\arctan(
	{\sqrt{\smash[b]{\Delta {d_{x,n-1}''}^2+\Delta {d_{y,n-1}''}^2}}/{\Delta d_{z,n-1}}}
	).
	\label{eq:delta_pp_def}
\end{equation}
where $\Delta d_{x,n-1}''$ and $\Delta d_{y,n-1}''$ denote the corresponding displacement components of $\Delta d_{n-1}''$ along the $x$- and $y$-axes in the considered coordinate system, respectively.

\begin{figure*}[!b]
	\hrulefill
	\begin{equation}
		\tilde{d}^2_{n-1}
		= {4	\left( 
			d^2_{n-1}\sin^2\varphi_{n-1}
			-\sqrt{\Delta d_{x,n-1}^2 +\Delta d_{y,n-1}^2}d_{n-1}\sin \varphi_{n-1}\cos(\theta_{n-1}+\delta_{n-1})\right)\sin^{2}
			\frac{\Delta d_{n-1}' \sin(\theta_{n-1}+\delta_{n-1})}
			{2d_{n-1} \sin \varphi_{n-1} }
		}.
		\label{eryuan1}
	\end{equation}
\end{figure*} 
Since \( \Delta d_{n-1}'' = \sqrt{\smash[b]{\Delta {d_{x,n-1}''}^2 + \Delta{d_{y,n-1}''}^2 + \Delta d_{z,n-1}^2}} \) and the value of \( \Delta d_{n-1}'' \) is already known from equation \eqref{qwerrewq}, we can derive the following
\begin{align}
	\Delta	{d_{x,n-1}''}^2+\Delta{d_{y,n-1}''}^2
	=\Delta d_{x,n-1}^2+\Delta d_{y,n-1}^2-\tilde d^2_{n-1}.
	\notag
\end{align}
Substituting the above expression into \eqref{eq:delta_pp_def} yields
\begin{align}
	\delta_{n-1}''
	=\arctan(
	{\sqrt{\smash[b]{\Delta d_{x,n-1}^2+\Delta d_{y,n-1}^2-\tilde d^2_{n-1}}}/{\Delta d_{z,n-1}}}
	).
	\notag
\end{align}

Therefore, by substituting the derived expressions of $\Delta d_{n-1}''$ and $\delta_{n-1}''$ into \eqref{Cs2} and rearranging the result, the final state evolution equations of $\theta_n$, $\varphi_n$, and $d_n$ can be obtained as
\begin{equation}\label{Fin22}
	\begin{cases}
		\theta_n=\theta_{n-1}+{\Delta d_{n-1}^{\prime}\sin(\theta_{n-1}+\delta_{n-1})}/{{d_{n-1}'}}+\omega_{\theta},\\
		\varphi_n=\varphi_{n-1}+{\Delta d_{n-1}^{\prime\prime}\sin(\varphi_{n-1}+{\delta}_{n-1}'')}/{d_{n-1}}+\omega_{\varphi},\\
		d_n=d_{n-1}-\Delta d_{n-1}^{\prime\prime}\cos(\varphi_{n-1}+{\delta}_{n-1}'')+\omega_{d},
	\end{cases}
\end{equation}
where $\omega_{\theta}\sim\mathcal{N}(0,\sigma_\theta^2)$, $\omega_{\varphi}\sim\mathcal{N}(0,\sigma_\varphi^2)$, and $\omega_{d}\sim\mathcal{N}(0,\sigma_d^2)$ denote the corresponding noise, respectively.
Here, $\sigma_\theta^2$, $\sigma_\varphi^2$, and $\sigma_d^2$ denote the approximation errors between the proposed evolution equations and the actual state, and are independent of time and measurement signals.
Finally, the state evolution of \(\mathbf{x}_n\) is obtained by combining the velocity update in \eqref{Fin11} and the angle-and-range update in \eqref{Fin22}.
For notational simplicity, the overall state evolution is written as \(\widehat{\mathbf{x}}_{n|n-1}=\mathbf{g}(\widehat{\mathbf{x}}_{n-1})\), with \(\mathbf{Q}_w\) denoting the process noise covariance.

\subsection{Measurement Equation}
\label{B-2}

In general, the UAV receives echoes from different parts of the USV rather than directly from the CR. Therefore, the USV is modeled as an extended target consisting of multiple scattering centers, as shown in Fig.~\ref{fig:usv_echo_processing}(a). 
In what follows, we first describe the parameter estimation process and then establish the measurement model for the CR. Finally, the EKF algorithm is summarized based on the derived state and measurement models.

\subsubsection{Echo Processing and Scatterer Detection}

\begin{figure}[!t]
	\centering
	\includegraphics[width=\columnwidth]{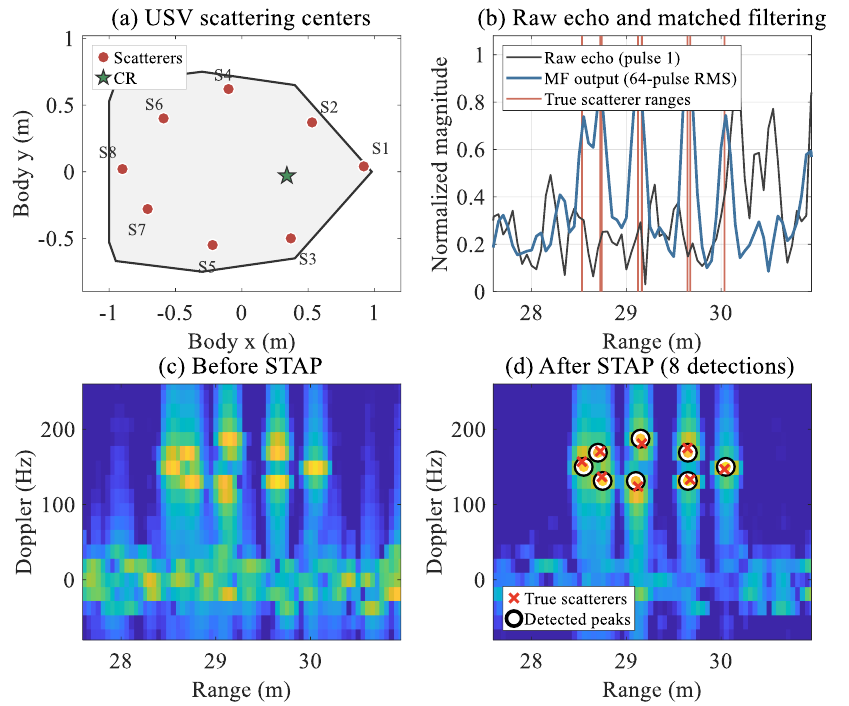}
	\caption{An illustrative example of USV echo
		processing: (a) USV scattering centers;
		(b) Raw received echo and matched-filtering
		output; (c) Range-Doppler map before STAP;
		(d) Range-Doppler map after STAP with
		detected USV scatterers.}
	\label{fig:usv_echo_processing}
\end{figure}


We consider a total of $L$ sensing pulses transmitted during each sensing epoch for echo processing.
Thus, the matched-filter outputs from all
$L$ pulses are stacked to form the space-time
observation vector
\begin{equation}
	\mathbf y_n^{\mathrm{MF}}(\tau)
	=
	\begin{bmatrix}
		(\tilde{\mathbf y}_{n,1}(\tau))^T
		& \cdots &
		(\tilde{\mathbf y}_{n,L}(\tau))^T
	\end{bmatrix}^{T},
	\label{eq:stap_input}
\end{equation}
where $\tilde{\mathbf y}_{n,l}(\tau)$ represents the matched-filter outputs for the $l$-th sensing pulse.
As shown in Fig.\ref{fig:usv_echo_processing}(c) and Fig.\ref{fig:usv_echo_processing}(d), the STAP method is applied to suppress sea clutter by exploiting spatial and temporal information. Specifically, the space-time covariance matrix of the sea-clutter signal can be expressed as
\begin{equation}
	\mathbf R_{c,n}(\tau)
	=
	\mathbb E\left[
	\mathbf i_n^{\mathrm{cl,MF}}(\tau)
	\bigl(\mathbf i_n^{\mathrm{cl,MF}}(\tau)\bigr)^H
	\right],
	\label{eq:clutter_covariance}
\end{equation}
where $\mathbf i_n^{\mathrm{cl,MF}}(\tau)$ represents the matched-filter output of the sea-clutter signal $\mathbf i_n^{\mathrm{cl}}(t)$. In practice, the matrix $\mathbf R_{c,n}(\tau)$ is difficult to obtain accurately. 
Therefore, we approximate the sea-clutter
covariance matrix as
$\mathbf R_{c,n}(\tau_{k,n})
\approx p\rho_n G P_{k,n}
\overline{\mathbf R}_{c,n}(\tau_{k,n})$,
under the following two assumptions.
First, the average sea-clutter power
remains constant during each sensing epoch.
Second, under the same sea-state level,
the average normalized space-time covariance
matrix is assumed to remain unchanged.
Here, $\rho_n$ is the fraction of the
sensing epoch allocated to wide-beam sensing,
$G$ is the effective processing gain for
a full sensing epoch, $P_{k,n}$ represents
the average clutter strength in the delay
cell of the $k$-th scatterer, and
$\overline{\mathbf R}_{c,n}(\tau_{k,n})$
describes the normalized space-time covariance,
satisfying
$\operatorname{tr}(
\overline{\mathbf R}_{c,n}(\tau_{k,n}))=1$.
Assuming that sea clutter and receiver noise are mutually uncorrelated, the corresponding clutter-plus-noise covariance matrix can be expressed as
\begin{equation}
	\mathbf R_{I,n}(\tau)
	=
	\mathbf R_{c,n}(\tau)
	+
	\sigma_0^2\mathbf I.
	\label{eq:stap_interference_covariance}
\end{equation}

In addition, the corresponding space-time
steering vector for each candidate search point
is given by
\begin{equation}
	\mathbf g_n(\theta,\varphi,\mu)
	=
	\mathbf d_L(\mu)
	\otimes
	\mathbf b(\theta,\varphi),
	\label{eq:stap_steering}
\end{equation}
where $\mathbf b(\theta,\varphi)$ is the
receive-array steering vector and
$\mathbf d_L(\mu)
=\frac{1}{\sqrt L}
[e^{j2\pi\mu t_1},\dots,e^{j2\pi\mu t_L}]^T$
describes the Doppler-induced phase changes
across the pulses.
Using the clutter-plus-noise covariance matrix and the corresponding space-time steering vector, the STAP weight at each candidate search point is then calculated as~\cite{s22072664}
\begin{equation}
	\begin{aligned}
		\mathbf w_n(\tau,\theta,\varphi,\mu)
		=
		\frac{
			\mathbf R_{I,n}^{-1}(\tau)
			\mathbf g_n(\theta,\varphi,\mu)
		}{
			\mathbf g_n^H(\theta,\varphi,\mu)
			\mathbf R_{I,n}^{-1}(\tau)
			\mathbf g_n(\theta,\varphi,\mu)
		}.
	\end{aligned}
	\label{eq:stap_weight}
\end{equation}
Therefore,
the STAP output is calculated as
\begin{equation}
	z_n(\tau,\theta,\varphi,\mu)
	=
	\mathbf w_n^H(\tau,\theta,\varphi,\mu)
	\mathbf y_n^{\mathrm{ST}}(\tau).
	\label{eq:stap_output}
\end{equation}
By searching for peaks in the output magnitude,
we obtain the delay, azimuth, elevation, and
Doppler estimates of the USV scatterers.
The detected scatterers form the set
$\mathcal K_n$, with
$K_n=|\mathcal K_n|$ denoting their number.

\subsubsection{CR Measurement Reconstruction}

For each detected scatterer $k\in\mathcal K_n$,
the delay, azimuth, elevation, and Doppler
estimates obtained from STAP are denoted by
$\widehat\tau_{k,n}$,
$\widehat\theta_{k,n}$,
$\widehat\varphi_{k,n}$, and
$\widehat\mu_{k,n}$, respectively.
The azimuth and elevation measurements are
modeled as
\begin{equation}
	\widehat\theta_{k,n}
	=
	\theta_{k,n}+z_{\theta_{k,n}},
	\quad
	\widehat\varphi_{k,n}
	=
	\varphi_{k,n}+z_{\varphi_{k,n}},
	\label{eq:scatterer_angle_meas}
\end{equation}
where $z_{\theta_{k,n}}$ and
$z_{\varphi_{k,n}}$ are the corresponding
measurement errors.
The delay and distance measurements are
\begin{equation}
	\widehat\tau_{k,n}
	=
	\frac{2d_{k,n}}{c}
	+z_{\tau_{k,n}},
	\qquad
	\widehat d_{k,n}
	=
	\frac{c\widehat\tau_{k,n}}{2},
	\label{eq:scatterer_delay_meas}
\end{equation}
where $d_{k,n}$ is the true distance,
$c$ is the speed of light, and
$z_{\tau_{k,n}}$ is the delay measurement
error.

Based on the above scatterer angle and distance measurements, we reconstruct the CR angle and range measurements $(\widehat\theta_n,\widehat\varphi_n,\widehat d_n)$ by first obtaining the estimated global Cartesian position of each retained scatterer as
\begin{equation}
	\widehat{\mathbf p}_{k,n}
	=
	\begin{bmatrix}
		\widehat d_{k,n}\sin\widehat\varphi_{k,n}
		\cos\widehat\theta_{k,n}
		\\
		\widehat d_{k,n}\sin\widehat\varphi_{k,n}
		\sin\widehat\theta_{k,n}
		\\
		H-\widehat d_{k,n}\cos\widehat\varphi_{k,n}
	\end{bmatrix}.
	\label{eq:scatterer_position_reconstruction}
\end{equation}

The USV centroid is then reconstructed by fusing all
retained scatterer-based 3D points as
\begin{equation}
	\widehat{\mathbf p}_{c,n}
	=
	\frac{1}{K_n}
	\sum_{k\in\mathcal K_n}
	\widehat{\mathbf p}_{k,n}.
	\label{eq:centroid_reconstruction}
\end{equation}

Since the EKF state is defined at the CR rather than at the USV centroid, the CR position is obtained using the known geometric offset,
$
\widehat{\mathbf r}_{n}
=
\widehat{\mathbf p}_{c,n}
+
\Delta\mathbf p_n
=
[
\widehat r_{x,n},
\widehat r_{y,n},
\widehat r_{z,n}
]^T,
$
where
$
\Delta\mathbf p_n
=
\mathbf T({\psi}_n)\Delta\mathbf p,
$
$\Delta\mathbf p$ denotes the fixed centroid-to-CR
offset in the USV body-fixed coordinate system\footnote{
Since the CR and scattering centers are rigidly attached to the USV, their relative positions are fixed and can be calibrated in advance. In the simulations, the known scattering-center offsets are compensated for before centroid estimation to reduce bias caused by missed detections.
	},
and $\mathbf T({\psi}_n)$ denotes the corresponding
yaw-rotation matrix that transforms the offset
into the global Cartesian coordinate system, i.e.,
\begin{equation}
	\mathbf T({\psi}_n)
	=
	\begin{bmatrix}
		\cos{\psi}_n & -\sin{\psi}_n & 0\\
		\sin{\psi}_n & \cos{\psi}_n & 0\\
		0 & 0 & 1
	\end{bmatrix}.
\end{equation}

Accordingly, the angle and range estimates of the CR are obtained from its reconstructed position as
\begin{equation}
	\begin{bmatrix}
		\widehat\theta_n
		\\
		\widehat\varphi_n
		\\
		\widehat d_n
	\end{bmatrix}
	=
	\begin{bmatrix}
		\operatorname{atan2}
		\left(
		\widehat r_{y,n},
		\widehat r_{x,n}
		\right)
		\\[1mm]
		\operatorname{atan2}
		\left(
		\sqrt{\widehat r_{x,n}^2+\widehat r_{y,n}^2},
		H-\widehat r_{z,n}
		\right)
		\\[1mm]
		\sqrt{
			\widehat r_{x,n}^2
			+
			\widehat r_{y,n}^2
			+
			(H-\widehat r_{z,n})^2
		}
	\end{bmatrix}.
	\label{eq:cr_angle_range_reconstruction}
\end{equation}

Next, we consider the velocity estimation of the CR.
Assuming that each scatterer has approximately the same velocity as the CR, the Doppler shift of the $k$-th scatterer measured by the UAV is modeled as
\begin{equation}
	\widehat\mu_{k,n}
	=
	\frac{2f_c}{c}
	\mathbf u_{k,n}^T
	[
		v_{x,n}\:\:\:
		v_{y,n}\:\:\:
		v_{z,n}
	]^T
	+z_{\mu_{k,n}},
	\label{eq:doppler_meas_k}
\end{equation}
where $f_c$ is the carrier frequency, $z_{\mu_{k,n}}$ is the Doppler measurement error, and $\mathbf u_{k,n}=[\sin\varphi_{k,n}\cos\theta_{k,n}, \sin\varphi_{k,n}\sin\theta_{k,n}, -\cos\varphi_{k,n}]^T$ is the LoS unit vector from the UAV to the scatterer.
By stacking the $K_n$ retained
Doppler measurements, the following linear
observation model is obtained
\begin{equation}
	\widehat{\boldsymbol{\mu}}_n
	=
	\mathbf A_n
	[v_{x,n},v_{y,n},v_{z,n}]^T
	+
	\mathbf z_{\mu,n},
	\label{eq:doppler_stack}
\end{equation}
where
\begin{equation} \notag
	\mathbf A_n
	=
	\frac{2f_c}{c}
	\begin{bmatrix}
		\mathbf u_{k_1,n}^T\\
		\mathbf u_{k_2,n}^T\\
		\vdots\\
		\mathbf u_{k_{K_n},n}^T
	\end{bmatrix}\:
	\widehat{\boldsymbol{\mu}}_n
	=
	\begin{bmatrix}
		\widehat{\mu}_{k_1,n}\\
		\widehat{\mu}_{k_2,n}\\
		\vdots\\
		\widehat{\mu}_{k_{K_n},n}
	\end{bmatrix}\:
	\mathbf z_{\mu,n}
	=
	\begin{bmatrix}
		z_{\mu_{k_1,n}}\\
		z_{\mu_{k_2,n}}\\
		\vdots\\
		z_{\mu_{k_{K_n},n}}
	\end{bmatrix}.
\end{equation}

Based on \eqref{eq:doppler_stack}, the velocity-related measurement components of the EKF are obtained by applying weighted least squares (WLS), i.e.,
\begin{equation}
		\widehat{\mathbf v}_n
		=
		[\widehat{v}_{x,n},
		\widehat{v}_{y,n},
		\widehat{v}_{z,n}]^T=
		(\widehat{\mathbf A}_n^T
		\mathbf Q_{\mu,n}^{-1}
		\widehat{\mathbf A}_n)^{-1}
		\widehat{\mathbf A}_n^T
		\mathbf Q_{\mu,n}^{-1}
		\widehat{\boldsymbol{\mu}}_n,
\notag
\end{equation}
where
$\mathbf Q_{\mu,n}
=
\mathrm{diag}(
\sigma_{\mu_{k_1,n}}^2,
\sigma_{\mu_{k_2,n}}^2,
\ldots,
\sigma_{\mu_{k_{K_n},n}}^2
)$
denotes the covariance matrix of the Doppler
measurement errors.
Combining the reconstructed angle-range quantities
$(\widehat{\theta}_n,\widehat{\varphi}_n,\widehat d_n)$
and the Doppler-based velocity estimate
$(\widehat v_{x,n},\widehat v_{y,n},\widehat v_{z,n})$,
the six-dimensional measurement vector is formed as
$
\mathbf y_n
\triangleq
[
\widehat{\theta}_n,\,
\widehat{\varphi}_n,\,
\widehat d_n,\,
\widehat v_{x,n},\,
\widehat v_{y,n},\,
\widehat v_{z,n}
]^T.
$
The EKF measurement model is approximated as
\begin{equation}
	\mathbf y_n
	=
	\mathbf x_n+\mathbf z_n,
	\label{eq:measurement_model_6d}
\end{equation}
where $\mathbf z_n\sim\mathcal N(\mathbf 0,\mathbf Q_{z,n})$
denotes the equivalent measurement noise.
Thus, based on the state evolution models
in \eqref{Fin11} and \eqref{Fin22}
and the measurement model in
\eqref{eq:measurement_model_6d},
the detailed EKF procedure is summarized
in Algorithm~\ref{alg:ekf}.

\subsubsection{Measurement Error Analysis}
\label{wuchafenxi}
The covariance matrix $\mathbf Q_{z,n}$ determined by the sea clutter and receiver noise plays a key role in the EKF algorithm. In this paper, we consider $\mathbf Q_{z,n}$ as a diagonal matrix, i.e.,
$\mathbf Q_{z,n}=\mathrm{diag}(
\bar\sigma_{\theta,n}^2,
\bar\sigma_{\varphi,n}^2,
\bar\sigma_{d,n}^2,
\bar\sigma_{v_x,n}^2,
\bar\sigma_{v_y,n}^2,
\bar\sigma_{v_z,n}^2
)$,
where the six diagonal elements denote the measurement error variances of the CR azimuth, elevation, range, and velocity components along the $x$-, $y$-, and $z$-axes, respectively.
It should be highlighted that these six measurement error variances originate from the associated variances of multiple scatters. 
Therefore, we model
the measurement error variances
$\sigma_{k,n,i}^2$ of individual detected
scatterers, where
$i\in\{\tau,\theta,\varphi,\mu\}$
corresponds to different measurement errors.
Based on $\sigma_{k,n,i}^2$, the
$\bar{\sigma}_{\theta,n}^2$,
$\bar{\sigma}_{\varphi,n}^2$, and
$\bar{\sigma}_{d,n}^2$ can be easily obtained via
first-order error propagation, while
$\bar{\sigma}_{v_x,n}^2$,
$\bar{\sigma}_{v_y,n}^2$, and
$\bar{\sigma}_{v_z,n}^2$ are obtained
via the WLS estimator.
For brevity, the derivations
are omitted, and the relevant procedures
can be found in the Appendix
of~\cite{9947033}.


Inspired by~\cite{9171304}, we adopt the inverse-SCNR approximation, under which these variances are inversely proportional to the corresponding SCNR.
For the $k$-th detected scatterer, the corresponding STAP weight is defined as
$\mathbf w_{k,n}$ as shown in \eqref{eq:stap_weight}.
Based on this weight and the clutter
covariance model, the residual sea-clutter
power after STAP is expressed as
\begin{equation}
		C_{k,n}^{\mathrm{out}}
		=
		\mathbf w_{k,n}^H
		\mathbf R_{c,n}(\tau_{k,n})
		\mathbf w_{k,n}\approx
		p\rho_nG P_{k,n}
		\xi_{k,n},
	\label{eq:stap_residual_power}
\end{equation}
where $\xi_{k,n}=\mathbf w_{k,n}^H\overline{\mathbf R}_{c,n}(\tau_{k,n})\mathbf w_{k,n}$ represents the residual clutter factor after STAP.
The useful echo power
after STAP is approximated as
$p\rho_nG|\kappa_n\beta_{k,n}|^2
|\varrho_{k,n}|^2$, where
$\varrho_{k,n}=\mathbf a_{N_{\mathrm{tx},n}}^H(\theta_{k,n},\varphi_{k,n})\mathbf f_n^{\mathrm{wide}}$
is the wide-beam response toward the
scatterer~\cite{9947033}.
Consequently, the SCNR corresponding to the $k$-th scatterer can be approximated as
\begin{equation}
	\begin{aligned}
		\mathrm{SCNR}_{k,n}
		&\approx
		\frac{
			p\rho_nG
			|\kappa_n\beta_{k,n}|^2
			|\varrho_{k,n}|^2
		}{
			p\rho_nG\xi_{k,n}P_{k,n}
			+
			\sigma_r^2
		}.
	\end{aligned}
	\label{eq:scnr_k}
\end{equation}
where $\sigma_r^2$ denotes the equivalent
receiver noise power.
Therefore, the variance of each
measurement error for scatterer $k$
can be expressed as
\begin{equation}
	\begin{aligned}
		\sigma_{k,n,i}^2
		&\approx
		\frac{a_i^2}{\mathrm{SCNR}_{k,n}}\\
		&\approx
		\frac{
			a_i^2\sigma_r^2
		}{
			p\rho_nG
			|\kappa_n\beta_{k,n}|^2
			|\varrho_{k,n}|^2
		}
		+
		\frac{
			a_i^2\xi_{k,n}P_{k,n}
		}{
			|\kappa_n\beta_{k,n}|^2
			|\varrho_{k,n}|^2
		}\\
		&\triangleq
		\frac{A_{k,n,i}}{\rho_n}
		+B_{k,n,i},
		\quad
		i\in\{\tau,\theta,\varphi,\mu\},
	\end{aligned}
	\label{eq:scatterer_meas_var}
\end{equation}
where $a_i$ is the corresponding error scaling factor.
\begin{algorithm}[!t]
	\caption{EKF-Based State Estimation}
	\label{alg:ekf}
	\begin{algorithmic}[1]
		
		\State \textbf{1) State Prediction:}
		\State \hspace{4em}
		$\widehat{\mathbf{x}}_{n|n-1}
		=
		\mathbf{g}(\widehat{\mathbf{x}}_{n-1})$
		
		\State \textbf{2) Linearization:}
		\State \hspace{4em}
		$\mathbf{G}_{n-1}
		=
		\left.
		\dfrac{\partial \mathbf{g}}
		{\partial \mathbf{x}}
		\right|_{\mathbf{x}=
			\widehat{\mathbf{x}}_{n-1}}$
		
		\State \textbf{3) Error Covariance Prediction:}
		\State \hspace{4em}
		${\mathbf{U}}_{n|n-1}
		=
		\mathbf{G}_{n-1}
		\mathbf{U}_{n-1}
		\mathbf{G}_{n-1}^{T}
		+
		\mathbf{Q}_w$
		
		\State \textbf{4) Measurement Covariance Construction:}
		\State \hspace{1em}
		Construct $\mathbf Q_{z,n}$ based on
		\eqref{eq:scatterer_meas_var} and the
		sea-clutter statistics
		$(\xi_{k,n}, P_{k,n})$,
		using first-order error propagation
		and the weighted least-squares covariance.
		
		\State \textbf{5) Kalman Gain:}
		\State \hspace{4em}
		$\mathbf{K}_n
		=
		\mathbf{U}_{n|n-1}
		\left(
		\mathbf{U}_{n|n-1}
		+
		\mathbf{Q}_{z,n}
		\right)^{-1}$
		
		\State \textbf{6) State Update:}
		\State \hspace{4em}
		$\widehat{\mathbf{x}}_n
		=
		\widehat{\mathbf{x}}_{n|n-1}
		+
		\mathbf{K}_n
		(
		\mathbf{y}_n
		-
		\widehat{\mathbf{x}}_{n|n-1}
		)$
		
		\State \textbf{7) Covariance Update:}
		\State \hspace{4em}
		$\mathbf{U}_n
		=
		(\mathbf I_6-\mathbf K_n)
		\mathbf U_{n|n-1}$
		
		\State \textbf{Output:}
		$\widehat{\mathbf{x}}_n$, $\mathbf{U}_n$
		
	\end{algorithmic}
\end{algorithm}

\textbf{\textit{Remark}}: Eq.~\eqref{eq:scatterer_meas_var} reveals two distinct contributions to the scatterer measurement error variance.
The first term decreases with increasing time allocation factor $\rho_n$, indicating that longer wide-beam sensing can improve CR state estimation accuracy, thereby reducing the probability of narrow-beam misalignment.
Conversely, this reduces the time available for narrow-beam communication, leading to a trade-off between sensing accuracy and communication rate.
The second term represents a sea-clutter-induced error floor that depends on the sea conditions and cannot be eliminated by increasing $\rho_n$.
Under severe sea conditions, more time may be allocated to wide-beam sensing to reduce the receiver-noise contribution and maintain reliable beam alignment.
In contrast, under mild sea conditions, the sensing time can be reduced to improve the communication rate. These observations motivate the adaptive time allocation optimization developed in the following subsection.

\subsection{Time Allocation Optimization}

In this subsection, we formulate a communication rate maximization problem and reformulate it into a tractable one-dimensional optimization problem by characterizing the beam alignment probability based on the EKF posterior angular uncertainty under sea clutter.

\subsubsection{Time Allocation Problem}

Under the proposed dual-mode scheme,
the overall communication rate at the
$n$-th time slot can be expressed as a
time-weighted sum of the wide- and
narrow-beam rates, leading to the
following optimization problem
\begin{equation}
	\begin{aligned}
		\mathrm{(P1):}\quad
		{\max}_{\rho_n}\quad
		&
		\rho_n R_n^{\mathrm{wide}}
		+
		(1-\rho_n) R_n^{\mathrm{narr}}
		\\
		\mathrm{s.t.}\quad
		&
		0<\rho_n\leq 1.
	\end{aligned}
	\label{eq:P1_time_allocation}
\end{equation}
where the communication rate $R_n^{\mathrm{wide}}$ in wide beam mode is given in \eqref{Wide}.
For the narrow-beam mode, the achievable rate
$R_n^{\mathrm{narr}}$ depends on whether
the beam is correctly aligned with the CR.
Denoting the beam alignment probability
by $P_{A,n}$, the average narrow-beam rate
can be approximated as
\begin{equation}
	R_n^{\mathrm{narr}}
	\approx
	P_{A,n}
	R_{\mathrm{align},n}^{\mathrm{narr}}
	+
	\left(1-P_{A,n}\right)
	R_{\mathrm{mis},n}^{\mathrm{narr}},
	\label{eq:R_narr_prob}
\end{equation}
where
$R_{\mathrm{align},n}^{\mathrm{narr}}$
and $R_{\mathrm{mis},n}^{\mathrm{narr}}$
denote the achievable rates under alignment
and misalignment, respectively.
For the convenience of analysis, we assume $R_{\mathrm{mis},n}^{\mathrm{narr}}\approx 0$ by ignoring the negligible received signal power in severe misalignment.
The aligned rate $R_{\mathrm{align},n}^{\mathrm{narr}}$ is given by
\begin{equation}
	R_{\mathrm{align},n}^{\mathrm{narr}}
	=
	\log_2
	\left(
	1+
	\frac{p
		N_xN_y
		|\widehat{\alpha}_{n}|^2
		\left|
		\widehat\varrho_{n}
		\right|^2
	}{
		\sigma_c^2
	}
	\right),
	\label{eq:R_narr_align}
\end{equation}
where
$
\widehat \varrho_{n}
=
{\mathbf a}_{N_{\mathrm{tx},n}}^H
(\theta_n,\varphi_n)
{\mathbf a}_{N_{\mathrm{tx},n}}
(\widehat{\theta}_n,\widehat{\varphi}_n)
$
denotes the beamforming alignment response
of the array.
Since the time allocation factor $\rho_n$ is determined before wide-beam sensing, while the estimates ($\widehat{\alpha}_n$ and $\widehat{\varrho}_n$) are still unavailable.
To address this issue, for a sufficiently small sampling interval, the estimated channel gain is approximated by its predicted channel gain, i.e., $|\widehat{\alpha}_n|^2 \approx|\widehat{\alpha}_{n|n-1}|^2$.
In addition, following~\cite{9947033}, we adopt the perfect beam alignment (PBA) approximation as a reasonable simplification, under which the beamforming gain loss is neglected, i.e., $|\widehat{\varrho}_{n|n-1}|^2 \approx|\widehat{\varrho}_n|^2 \approx 1$.
Therefore, the problem $\mathrm{(P1)}$ can be reformulated as
\begin{equation}
	\begin{aligned}
		\mathrm{(P2): \:}
		&{\max}_{\rho_n}\:
		\rho_n
		\log_2
		\left(
		1+
		\frac{
			p N_{\mathrm{tx},n}
			|\widehat{\alpha}_{n|n-1}|^2
		}{
			\sigma_c^2
		}
		\right)
		\\
		+
		(1&-\rho_n)
		P_{A,n}
		\log_2
		\left(
		1+
		\frac{
			p N_xN_y
			|\widehat{\alpha}_{n|n-1}|^2
		}{
			\sigma_c^2
		}
		\right)
		\\
		\mathrm{s.t.}\quad
		&
		0<\rho_n\leq 1,
	\end{aligned}
	\label{eq:P2_time_allocation}
\end{equation}
where $P_{A,n}$ and $N_{\mathrm{tx},n}$ need to be determined.
Since $\rho_n$ is determined before wide-beam sensing, the actual posterior angular uncertainty is unavailable, making $P_{A,n}$ difficult to evaluate directly.
Therefore, we approximate this uncertainty
using the expected EKF posterior error
covariance at each epoch.

\begin{algorithm}[!t]
	\caption{Beam Tracking Scheme}
	\label{alg:ISAC}
	\begin{algorithmic}[1]
		\State \textbf{Input:} $\hat{\mathbf{x}}_{0}$,  $T$, $N$, $N_x$, $N_y$, $\mathbf{Q}_w$.
		\While{$n \leq N$}
		\State  Update the number of antennas according to (\ref{eq:antenna_update_wb1}).
		\State Solve problem $(\mathrm{P2})$ to obtain $\rho_n$.
		
		\State \textbf{Wide Beam Mode:} $t\in[(n-1)\Delta T,(n-1+\rho_n)\Delta T]$
		
		\State  Update the state estimate $\hat{\mathbf{x}}_{n}$ using Algorithm~\ref{alg:ekf}.

		\State \textbf{Narrow Beam Mode:} $t\in[(n-1+\rho_n)\Delta T,n\Delta T]$
		\State Set ${\mathbf{f}}^{\mathrm{narr}}_n={\mathbf{a}}_{N_{\mathrm{tx},n}}(\widehat{\theta}_n,\widehat{\varphi}_n)$ and update $n \gets n+1$.
		\EndWhile
	\end{algorithmic}
\end{algorithm}

\subsubsection{Beam Alignment Probability}
By recalling the EKF algorithm~\ref{alg:ekf}, $\mathbf U_n$
denotes the posterior error covariance matrix.
Under the Gaussian approximation
with covariance $\mathbf U_n$,
the CR angular estimation errors
are modeled as
\begin{equation}
	\begin{bmatrix}
		\widehat{\theta}_n-\theta_n
		\\
		\widehat{\varphi}_n-\varphi_n
	\end{bmatrix}
	\sim
	\mathcal N
	\left(
	\mathbf 0,
	\begin{bmatrix}
		[\mathbf U_n]_{1,1}
		&
		[\mathbf U_n]_{1,2}
		\\
		[\mathbf U_n]_{2,1}
		&
		[\mathbf U_n]_{2,2}
	\end{bmatrix}
	\right).
	\label{eq:angular_error_gaussian}
\end{equation}

The narrow beam is regarded as successfully
aligned when the angular estimation errors
remain within the corresponding half-beamwidths,
namely $|\widehat{\theta}_n-\theta_n|
\leq\eta_{\theta,n}$ and
$|\widehat{\varphi}_n-\varphi_n|
\leq\eta_{\varphi,n}$.
Based on the predicted angular direction,
the corresponding half-beamwidths of
the UPA narrow beam can be approximated as
\begin{equation}
	\eta_{\theta,n}
	=
	\frac{0.89}
	{N_x
		\sin\widehat{\varphi}_{n|n-1}
		\cos\widehat{\theta}_{n|n-1}},
	\:
	\eta_{\varphi,n}
	=
	\frac{0.89}
	{N_y
		\sin\widehat{\varphi}_{n|n-1}}.
	\notag
\end{equation}

Therefore, the beam alignment probability is given by
\begin{equation}
	\begin{aligned}
		P_{A,n}
		&=
		\Pr
		\left(
		|\widehat\theta_n-\theta_n|
		\leq
		\eta_{\theta,n},
		|\widehat\varphi_n-\varphi_n|
		\leq
		\eta_{\varphi,n}
		\right)
		\\
		&\approx
		\operatorname{erf}
		\left(
		\frac{\eta_{\theta,n}}
		{\sqrt{2[\mathbf U_n]_{1,1}}}
		\right)
		\operatorname{erf}
		\left(
		\frac{\eta_{\varphi,n}}
		{\sqrt{2[\mathbf U_n]_{2,2}}}
		\right).
	\end{aligned}
	\label{eq:PA_rho_erf_angular}
\end{equation}

It is worth noting that sea-clutter power
affects the measurement covariance $\mathbf Q_{z,n}$,
which in turn changes the beam alignment
probability $P_{A,n}$ and indirectly influences
the time allocation problem.

\subsubsection{Transmit Subarray Size Selection}
\label{subsec:antenna_update_wb}

For a half-wavelength UPA, the half-power
beamwidths along the two angular dimensions
can be approximated as~\cite{11535301}
\begin{equation}
	\theta_{\mathrm{BW}}^{(x)} \approx
	\frac{1.78}
	{N_{x,n}\sin\widehat\varphi_{n|n-1}
		\cos\widehat\theta_{n|n-1}},
	\theta_{\mathrm{BW}}^{(y)} \approx
	\frac{1.78}
	{N_{y,n}\sin\widehat\varphi_{n|n-1}}.
	\notag
\end{equation}

The corresponding effective beam footprint
on the horizontal plane is approximated by
\begin{equation}
	\Delta_x =
	2\widehat d_{n|n-1}
	\tan\left(\frac{\theta_{\mathrm{BW}}^{(x)}}{2}\right),
	\Delta_y =
	2\widehat d_{n|n-1}
	\tan\left(\frac{\theta_{\mathrm{BW}}^{(y)}}{2}\right).
	\notag
\end{equation}

To fully cover the extended USV, the beam
footprint should satisfy
$\Delta_x\geq\Delta D$ and
$\Delta_y\geq\Delta D$, where $\Delta D$
denotes the maximum horizontal projection
length of the USV.
Accordingly, the subarray sizes are
determined as
\begin{equation}
	\begin{aligned}
		N_{x,n}
		&\approx
		\frac{0.89}{
			\sin\widehat\varphi_{n|n-1}
			\cos\widehat\theta_{n|n-1}
			\arctan\left(
			\frac{\Delta D}{2\widehat d_{n|n-1}}
			\right)
		},\\
		N_{y,n}
		&\approx
		\frac{0.89}{
			\sin\widehat\varphi_{n|n-1}
			\arctan\left(
			\frac{\Delta D}{2\widehat d_{n|n-1}}
			\right)
		}.
	\end{aligned}
	\label{eq:antenna_update_wb1}
\end{equation}
The resulting values are rounded down to integers and limited by the maximum array dimensions $N_x$ and $N_y$, respectively.
Finally, using the subarray size determined
by \eqref{eq:antenna_update_wb1} and
substituting \eqref{eq:PA_rho_erf_angular}
into \eqref{eq:P2_time_allocation},
problem $\mathrm{(P2)}$ can be solved
numerically via a one-dimensional grid
search over $0<\rho_n\leq 1$.
The overall beam tracking procedure is
summarized in Algorithm~\ref{alg:ISAC}.

\section{Simulations}


In this section, we validate the effectiveness and
robustness of the proposed sea-air beam tracking framework. The main simulation parameters are summarized in Table~\ref{tab:sim_params}.
We first generate realistic CR trajectories under sea-induced disturbances. Subsequently, the proposed framework is evaluated in terms of beam tracking and communication performance, generalization under unseen sea states, the range-dependent extended target effect, and the impact of sea-clutter power.

\begin{table}[!t]
	\centering
	\caption{Simulation parameters}
	\label{tab:sim_params}
	\footnotesize
	\setlength{\tabcolsep}{14pt}
	\renewcommand{\arraystretch}{1.0}
	\begin{tabular}{@{}c c c c@{}}
		\toprule
		Parameter & Value & Parameter & Value \\
		\midrule
		
		$T$ & $80~\mathrm{s}$
		& $\Delta T$ & $0.1~\mathrm{s}$ \\
		
		$G$ & $128$
		& $H$ & $100~\mathrm{m}$ \\
		
		$p$ & $1~\mathrm{W}$
		& $f_c$ & $10~\mathrm{GHz}$ \\
		
		$\boldsymbol{\tau}_n$
		& $[15,\,85,\,0.5]^T$
		& $\alpha_{\mathrm{ref}}$ & $1$ \\
		
		$\sigma_c^2$ & $10^{-6}$
		& $\sigma_r^2$ & $10^{-6}$ \\
		
		$K$ & $6$
		& $N_{\mathrm{rx}}$ & $324$ \\
		
		$N_x,\,N_y$ & $18,\,18$
		& $\Delta D$ & $10~\mathrm{m}$ \\
		
		$H_s$ & $0.5~\mathrm{m}$
		& $\omega_p^e$ & $1.35~\mathrm{s}$ \\
		
		$\gamma$ & $5.0$
		& $N_\omega$ & $300$ \\

		$P_{k,n}$ & $0.01$
		& $\xi_{k,n}$ & $100$ \\
		
		
		$X_{k,n}^{\mathrm{sen}}$
		& $\mathcal{N}(0,1.5^2)$
		& $K_{k,n}^{\mathrm{sen}}$ & $8~\mathrm{dB}$ \\
		
		$A_n^{\mathrm{com}}$
		& $\mathcal{N}(0,1^2)$
		& $K^{\mathrm{com}}$ & $10~\mathrm{dB}$ \\
		
		$\sigma_\theta$ & $0.05^\circ$
		& $\sigma_\varphi$ & $0.05^\circ$ \\
		
		$\sigma_d$ & $0.01~\mathrm{m}$
		& $\sigma_x$ & $0.01~\mathrm{m/s}$ \\
		
		$\sigma_y$ & $0.01~\mathrm{m/s}$
		& $\sigma_z$ & $0.01~\mathrm{m/s}$ \\
		
		\bottomrule
	\end{tabular}
\end{table}

\subsection{CR Motion Generation}
\begin{figure}[!t]
	\centering
	\includegraphics[width=6.5cm]{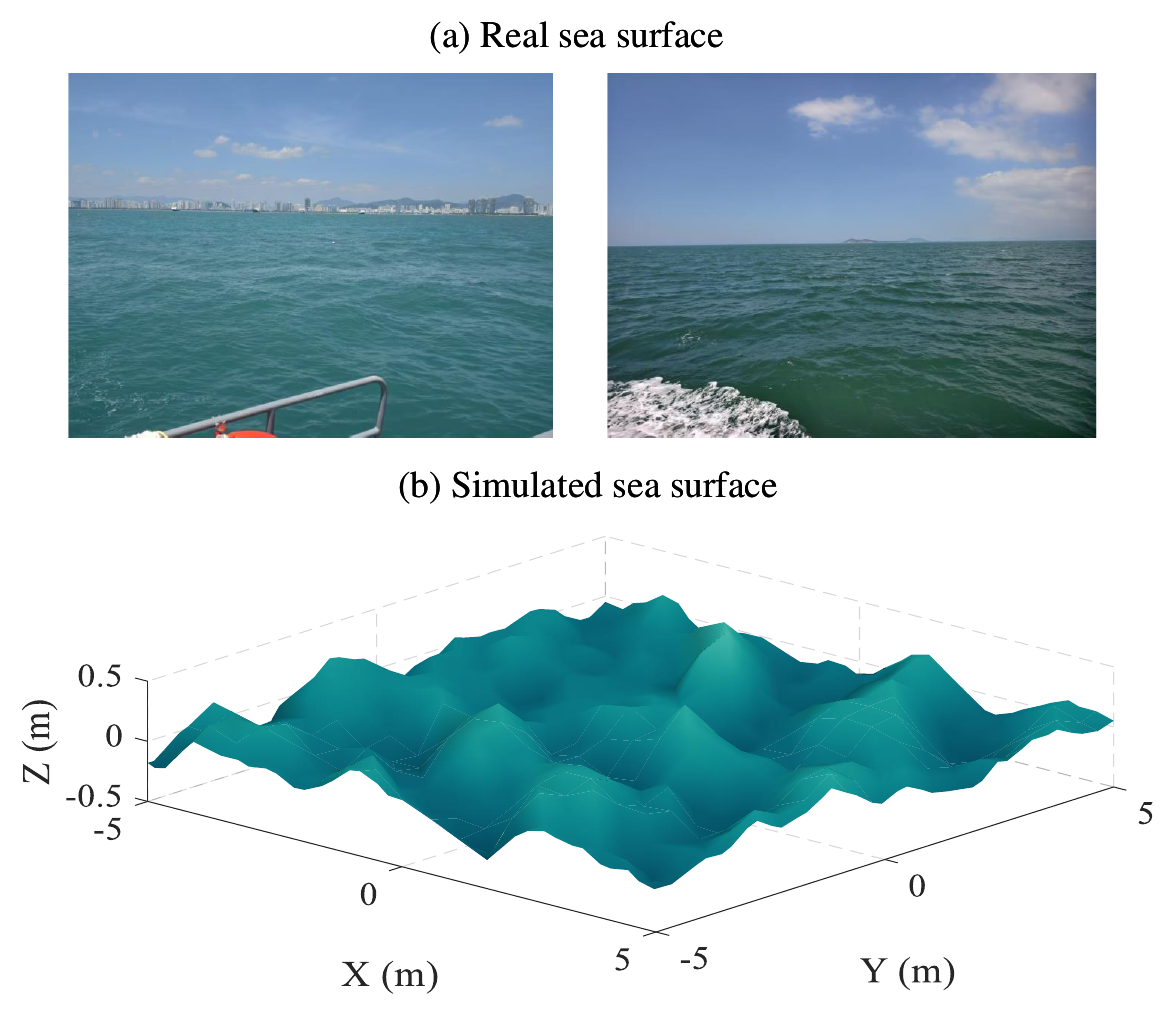}
	\caption{Real and simulated sea surfaces:
		(a) Photograph taken near Sanya, Hainan,
		China (18.2340$^\circ$N, 109.4624$^\circ$E);
		(b) Sea-surface elevation generated
		using the JONSWAP spectrum.}
	\label{fig:sea_surface}
\end{figure}

\begin{figure}[!t]
	\centering
	\includegraphics[width=6.5cm]{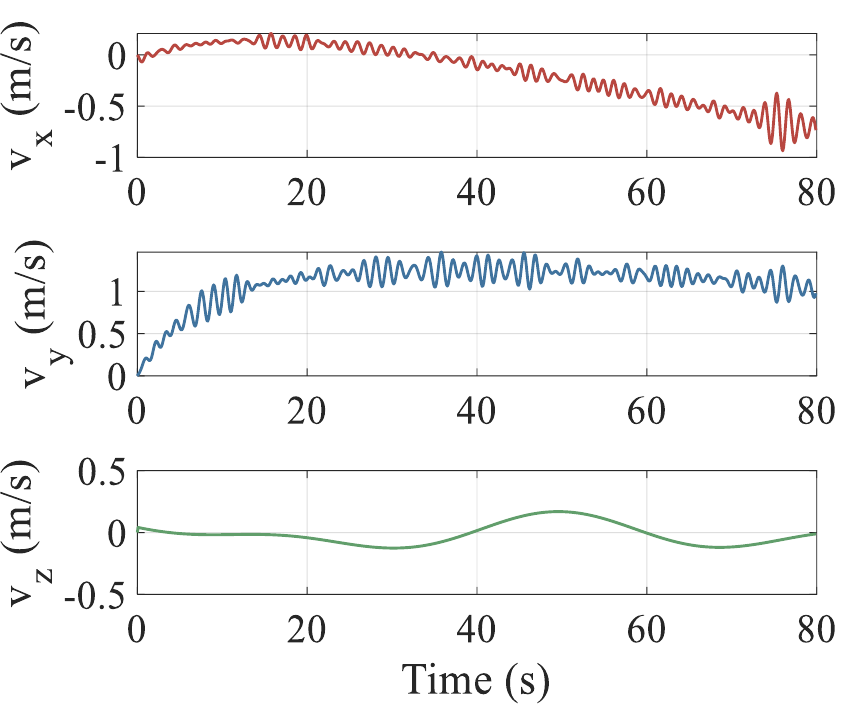}
	\caption{CR speed versus time.}
	\label{test2:sudu}
\end{figure}

To generate realistic CR velocities under maritime navigation, we construct a time varying sea induced load model based on \eqref{hailang}. 
Since the wave loads depend on local wave parameters and the USV hydrodynamic mapping, an irregular sea surface is synthesized using the Joint North Sea Wave Project (JONSWAP) spectrum~\cite{8488539}. 
The spectrum characterizes the wave-energy distribution over the frequency domain and is expressed as
\begin{equation} \notag
	S(\omega)
	=
	\left(1 - 0.287\ln\gamma\right)
	\frac{5}{16}
	\frac{H_s^2 \omega_p^4}{\omega^5}
	\exp\left(-\frac{5}{4}\frac{\omega_p^4}{\omega^4}\right)
	\gamma^{\alpha(\omega)},
\end{equation}
where $H_s$ and $\gamma$ denote the significant wave height and peak enhancement factor, respectively, and $\omega_p=2\pi/\omega_p^e$, with $\omega_p^e$ denoting the peak period. 
The exponent term $\alpha(\omega)$ is
$
	\alpha(\omega)
	=
	\exp(
	-\frac{(\omega-\omega_p)^2}{2\omega_p^2\sigma_\omega^2}
	),
$
where $\sigma_\omega$ is the spectral width parameter.
The frequency range is discretized into $N_\omega$ components with interval $\Delta\omega$, and the amplitude of the $m$-th component is obtained as $a_m=\sqrt{2S(\omega_m)\Delta\omega}$. 
The stochastic wave elevation is then synthesized as
\begin{equation}
	\begin{aligned}
		\eta_\Delta(x,y,t)
		=
	&	\sum_{m=1}^{N_\omega}
		a_m
		\cos(
		\omega_m t
		\\&	-
		k_m(x\cos\beta_m+y\sin\beta_m)
		+
		\varepsilon_m
		),
	\end{aligned}
\end{equation}
where $(x,y)$ is the horizontal position, $k_m$ and $\beta_m$ are the wave number and propagation direction of the $m$-th component, respectively, and $\varepsilon_m\sim\mathcal U(0,2\pi)$ is the random phase. 
As shown in Fig.~\ref{fig:sea_surface}, the sea surface
generated using the JONSWAP spectrum
exhibits irregular wave patterns similar
to those observed in the real maritime scene.
One possible approach is to evaluate the local wave elevation at the current CR position and extract the dominant wave component to update the wave-load parameters in \eqref{hailang}.
Alternatively, drawing on the spectral-response modeling approach in~\cite{5338560}, the time-domain wave-induced forces can be synthesized by combining the available wave spectrum with DoF-specific force response functions.
For computational simplicity, this approach is adopted in the simulations.
Finally, the resulting wave-excitation loads are incorporated into the USV dynamic model with the self-propulsion parameters in Table~\ref{tab:sim_params}, yielding the time-varying CR velocity components shown in Fig.~\ref{test2:sudu}. 
These generated velocities are used as the ground-truth CR velocities for subsequent evaluation.

\subsection{Beam Tracking and Communication Performance}

\begin{table*}[t]
	\centering
	\caption{\textsc{Performance comparison of the proposed scheme and baseline methods}}
	\label{tab:comparison}
	\resizebox{\textwidth}{!}{
		\begin{tabular}{c c c c c c c c}
			\toprule
			Method 
			& $\theta$ RMSE ($^\circ$) 
			& $\phi$ RMSE ($^\circ$) 
			& $d$ RMSE (m) 
			& $v_x$ RMSE (m/s) 
			& $v_y$ RMSE (m/s) 
			& $v_z$ RMSE (m/s) 
			& Throughput (bits) \\
			\midrule
			
			Proposed  
			& \textbf{0.0551} 
			& \textbf{0.0453} 
			& 0.4691 
			& 0.0270 
			& \textbf{0.0274} 
			& 0.0117 
			& \textbf{1159.73} \\
			
			KalmanNet  
			& 0.3920 
			& 0.0812 
			& \textbf{0.3166} 
			& \textbf{0.0220} 
			& 0.0281 
			& \textbf{0.0084} 
			& 1159.49 \\
			
			CM-EKF  
			& 30.3985 
			& 11.3835 
			& 12.6886 
			& 0.3697 
			& 0.8670 
			& 0.1560 
			& 373.00 \\
			
			Point-Target  
			& 9.7269 
			& 13.7886 
			& 18.6680 
			& 0.3604 
			& 1.0881 
			& 0.1724 
			& 311.42 \\
			
			\bottomrule
		\end{tabular}
	}
\end{table*}
\begin{figure*}[t]
	\centering
	\subfloat[\normalfont \label{fig8a} State estimation performance.]{
		\includegraphics[width=8.6cm]{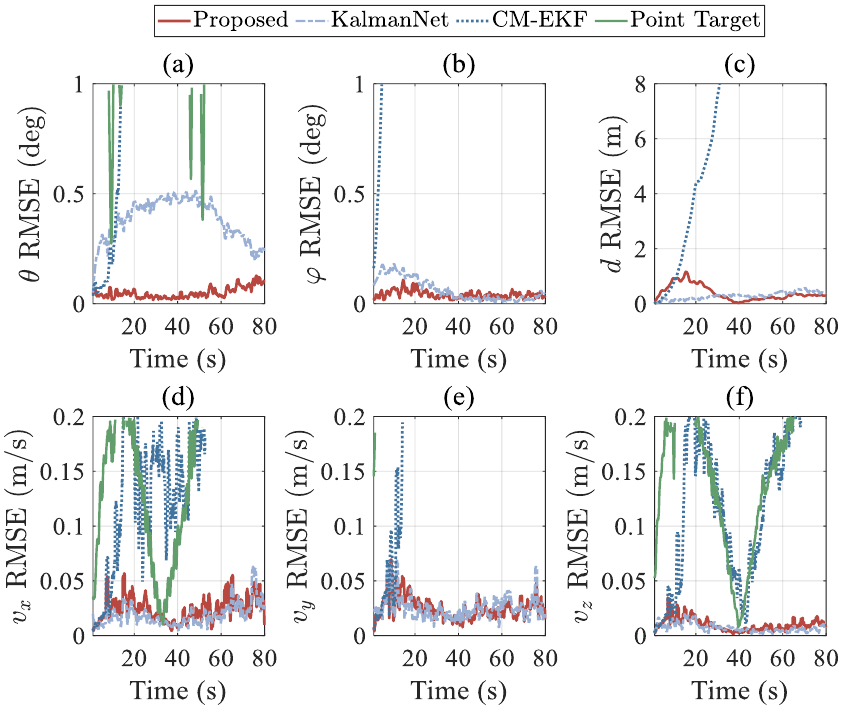}}
	\subfloat[\normalfont \label{fig8b} Communication performance.]{
		\includegraphics[width=8.6cm]{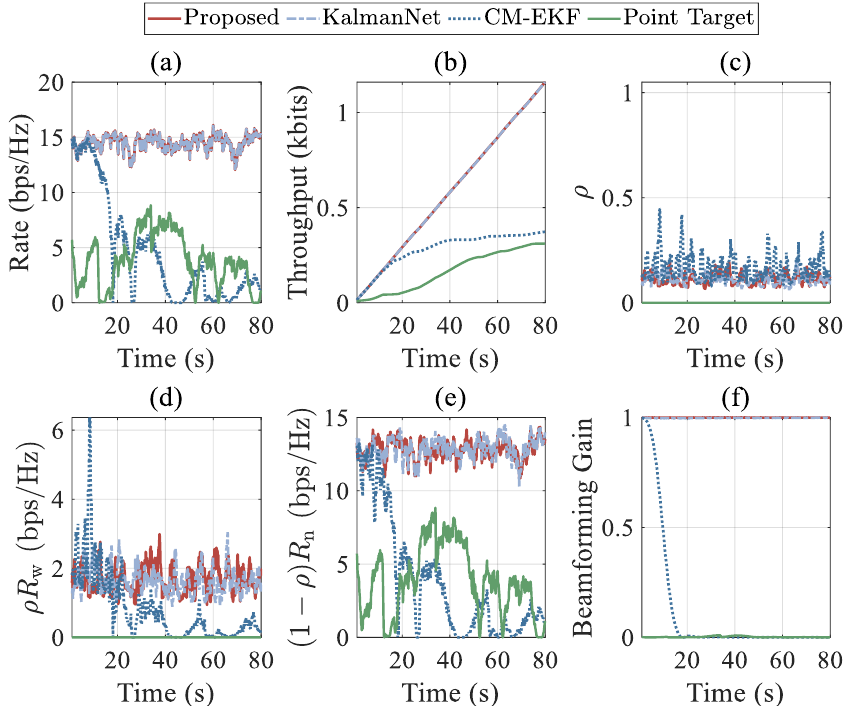}}
	\caption{Comparison of state estimation and communication performance for different schemes.}
	\label{fig8}
\end{figure*}

To demonstrate the advantages of the proposed beam tracking scheme in Algorithm~\ref{alg:ISAC}, the following three baseline schemes are considered.
\begin{itemize}

\item
\textit{KalmanNet Scheme} \cite{9733186}:
The USV is modeled as a 2D extended target, with conventional state prediction and a KalmanNet trained to learn the Kalman gain for measurement-based state correction. The proposed adaptive beam transmission and time allocation strategies are retained.

	\item
	\textit{Conventional-Motion EKF (CM-EKF) Scheme} \cite{9947033}:
	The USV is modeled as a 2D extended target and tracked using a
	conventional EKF, while adaptive beam selection is performed based
	on the estimated state and tracking uncertainty.
	
	\item
	\textit{Point Target Scheme} \cite{9171304}:
	The USV is modeled as a point target and tracked using a conventional
	EKF, while narrow-beam transmission is performed using the full
	antenna array.
	
\end{itemize}

Fig.~\ref{fig8a} and Table \ref{tab:comparison} compare the tracking performance of the four schemes. 
The proposed scheme and KalmanNet show similar overall tracking accuracy, while the proposed scheme achieves smaller angular errors. 
This is mainly because the proposed sea-air
state evolution model explicitly describes the motion variation caused
by sea-induced disturbances, which improves the prediction of the
azimuth and elevation states. KalmanNet achieves slightly smaller
errors in range and velocity because its learned correction can
compensate for part of the model mismatch under the considered
operating condition. In contrast, CM-EKF relies on a conventional
motion model and therefore cannot fully capture the sea-induced USV
dynamics. The point target scheme further ignores the spatial extent of
the USV, which introduces additional sensing and tracking errors.
These results show that sea-air motion prediction and extended target sensing are both important for accurate maritime tracking.

Fig.~\ref{fig8b} shows the corresponding communication performance.
The proposed scheme and KalmanNet maintain high beamforming gains and achieve almost the same accumulated throughput. 
Although KalmanNet achieves good performance after offline training, the proposed scheme does not require neural-network training. 
Instead, it improves the EKF by using the sea-air state evolution model together with adaptive sensing and beam transmission. 
As a result, the proposed framework can achieve performance close to a well-trained KalmanNet while avoiding the extra training cost and dependence on training data.
This result shows that a well-designed physical model can help a conventional EKF achieve performance close to that of a data-driven estimator.

\begin{figure*}[t]
	\centering
	\subfloat[\normalfont \label{fig101} Throughput versus significant wave height.]{
		\includegraphics[width=4.2cm]{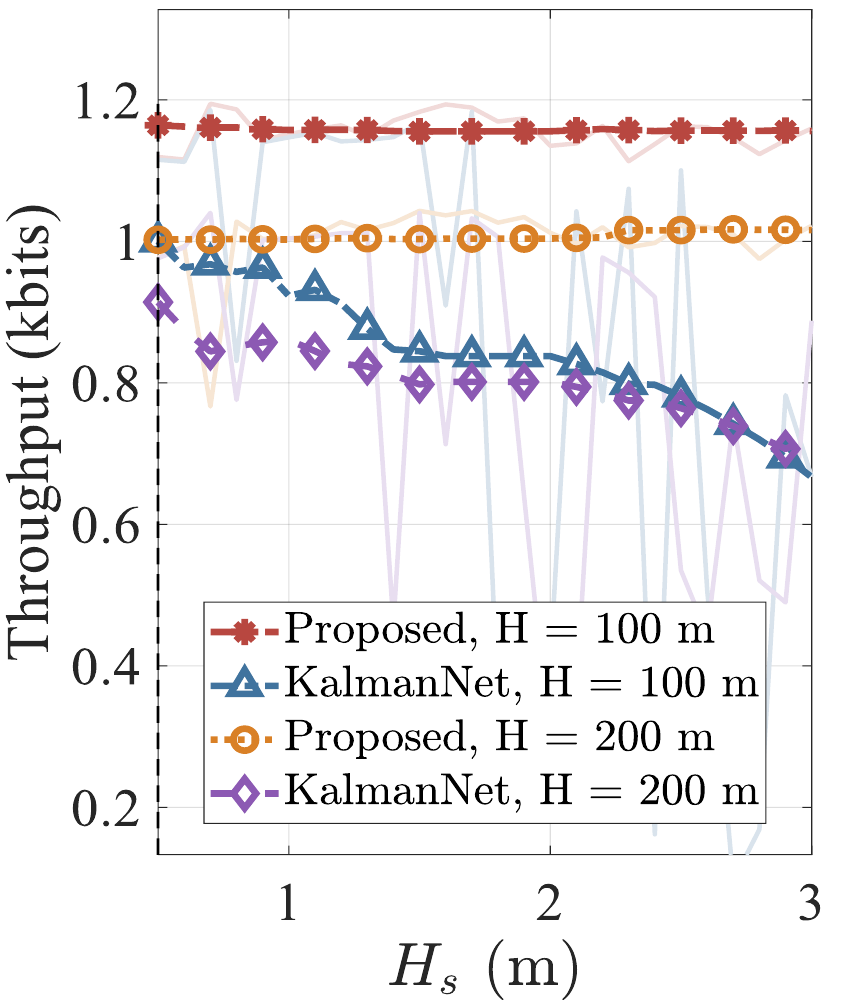}}
	\subfloat[\normalfont \label{fig10} Throughput versus UAV altitude.]{
		\includegraphics[width=4.2cm]{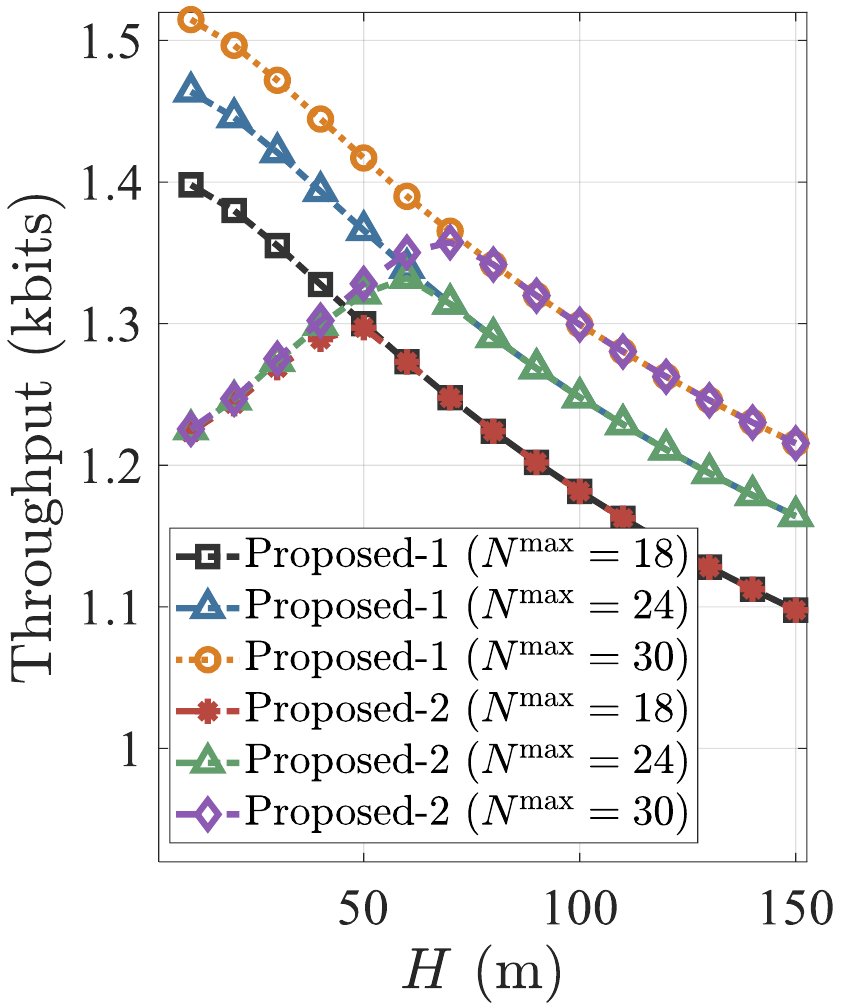}}
			\subfloat[\normalfont \label{fig11} Throughput versus sea clutter power.]{
			\includegraphics[width=4.2cm]{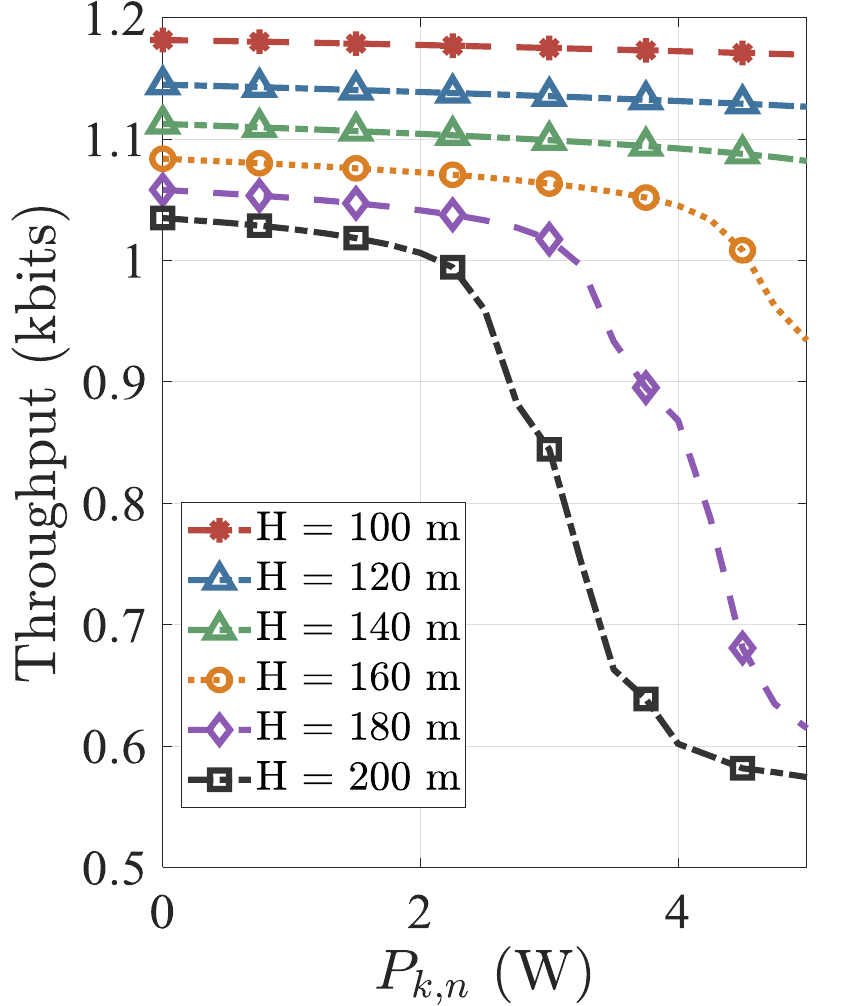}}
				\subfloat[\normalfont \label{fig12} Throughput versus sea clutter power.]{
				\includegraphics[width=4.2cm]{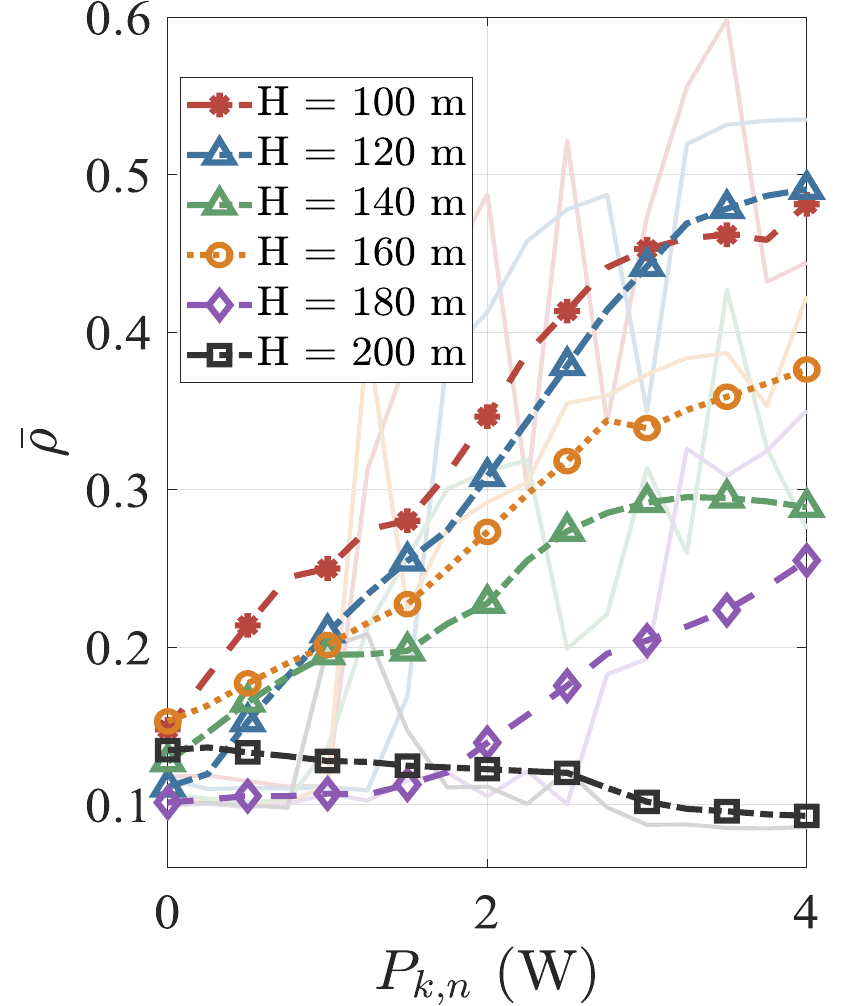}}
\caption{Impact of sea conditions and UAV altitude
	on communication throughput. To better illustrate the overall trends,
	the curves in Figs.~\ref{fig101} and
	\ref{fig12} are smoothed using a moving
	average, while the original curves are
	shown in lighter colors.}
\end{figure*}

\subsection{Generalization Performance}

To evaluate the generalization performance of the proposed framework under unseen sea conditions, the peak period $\omega_p^e$ is changed from $1.35$ s to $2.35$ s, and the sampling interval is increased from $0.1$ s to $0.2$ s for both schemes. 
As shown in Fig.~\ref{fig101}, the proposed scheme maintains nearly stable throughput as $H_s$ increases from $0.5$ m to $3$ m at both $H=100$ m and $H=200$ m. 
In contrast, the throughput of KalmanNet gradually decreases as the sea conditions become more severe, with a more pronounced performance loss at $H=200$ m. 
This is because the proposed scheme directly updates the state estimate using the physical model and current measurements, while KalmanNet relies on a learned mapping that becomes less accurate when the sea conditions differ from its training data. 
These results demonstrate that the proposed scheme provides better generalization to unseen sea conditions without requiring any training process.


\subsection{Extended-Target Effect}

To better illustrate the extended target
effect, particularly
the behavior resembling a point target
at long distances, we consider the
following two schemes.
\begin{itemize}
	\item
	\textit{Proposed-1 Scheme}:
	This scheme employs the proposed EKF
	with optimized time allocation $\rho_n$
	between wide- and narrow-beam modes.
	
	\item
	\textit{Proposed-2 Scheme}:
	This scheme employs the proposed EKF
	with wide-beam transmission only
	($\rho_n=1$), where the transmit subarray
	size is updated at each epoch according
	to \eqref{eq:antenna_update_wb1}.
\end{itemize}

Fig.~\ref{fig10} shows the throughput versus the UAV altitude $H$ under different maximum array sizes $N^{\max}=N_x=N_y$. For the proposed-1 scheme, the throughput decreases with $H$ due to the increased air-sea propagation distance and the resulting path loss. In contrast, for the proposed-2 scheme, the throughput first increases and then decreases. At low altitudes, the USV occupies a large angular region, requiring a wide beam and thus limiting the number of activated antennas and the achievable beamforming gain. As $H$ increases, the angular spread of the USV decreases, allowing more antennas to be selected and improving the beamforming gain. When the propagation distance becomes sufficiently large, however, path loss dominates and the throughput decreases. Moreover, a larger $N^{\max}$ provides higher array gain and thus improves the throughput. 
These results indicate that the proposed scheme is particularly effective at short distances, where the extended target effect is pronounced. As the UAV-USV distance increases, the extended target effect gradually diminishes and the USV approaches point target behavior, leading to a smaller performance advantage over conventional point-target beamforming schemes.


In addition, the performance gap between
the proposed-1 and proposed-2 schemes
mainly arises from the reduced number of
active transmit antennas required for full
USV coverage at short ranges.
Continuous wide-beam transmission therefore
results in considerable array gain loss.
By optimizing $\rho_n$, the proposed
time allocation scheme reduces this loss
through narrow-beam communication while
maintaining sufficient wide-beam sensing,
thereby achieving a better trade-off
between sensing and communication under
strong extended-target effects.

\subsection{Impact of Sea Clutter Power}

Fig.~\ref{fig11} shows the throughput variation with the sea clutter power \(P_{k,n}\) under different UAV altitudes. A larger \(P_{k,n}\) indicates stronger sea clutter and more severe sea conditions. It can be observed that the throughput decreases as \(P_{k,n}\) increases for all considered altitudes, since stronger sea clutter reduces the effective sensing SCNR and degrades the CR state estimation accuracy. As a result, the narrow beam is more likely to deviate from the CR, causing beam misalignment and lower achievable throughput. 
These results highlight the importance of explicitly accounting for sea-clutter-induced measurement uncertainty in the proposed method.

To further investigate how the proposed framework adapts to varying sea clutter conditions, we define the average time allocation factor as $\bar{\rho}=\frac{1}{N}\sum_{n=1}^{N}\rho_n$.
Fig.~\ref{fig12} shows how \(\bar{\rho}\) varies with the sea clutter power \(P_{k,n}\) under different UAV altitudes with \(\xi_{k,n}=1000\).
It can be seen that \(\bar{\rho}\) increases as \(P_{k,n}\) grows. This is because stronger sea clutter reduces the sensing SCNR and increases the estimation error. Therefore, more sensing time is allocated to the wide beam mode to improve accuracy.
In addition, a lower UAV altitude generally requires a larger
$\bar{\rho}$, as the extended-target effect of the USV becomes more
pronounced and the distributed scatterer echoes are more susceptible
to sea clutter. 
Overall, these results illustrate how the sensing time allocation adapts to different sea clutter conditions and further demonstrate the effectiveness of the proposed time allocation scheme.

\section{Conclusion and Future Work}
\label{Sec:6}
This paper proposed a predictive beam tracking framework for sea-air communication. By modeling the USV as an extended target with multiple distributed scatterers, a sea-air state-space model was established to capture the CR state evolution under the coupled surge, sway, and yaw motions induced by sea disturbances. 
Based on this model, an EKF-based predictive tracking method incorporating sea clutter effects was developed, and a time allocation strategy was further optimized to balance sensing accuracy and communication performance. 
Simulation results show that
(i) the range-dependent applicability
boundary of the extended-target model
is identified,
(ii) the impact of sea clutter intensity
on wide-beam sensing time allocation
is characterized, and
(iii) the proposed framework achieves
reliable beam tracking and more stable throughput under unseen
sea conditions compared with
learning-based benchmarks.

Future work will extend the proposed framework
to more realistic maritime scenarios by
incorporating UAV mobility and the roll and
pitch motions of the USV. In addition,
uncertainties in hydrodynamic parameters, and
imperfect sea-state information will be considered
to improve the robustness of beam tracking. 
\appendices
\section{USV Dynamic Matrices}
\label{app:usv_matrices}
The mass-inertia matrix $\mathbf M$ in
\eqref{3DofUSV} is expressed as
\begin{equation}
	\mathbf M =
	\begin{bmatrix}
		m_{11} & 0 & 0\\
		0 & m_{22} & m_{23}\\
		0 & m_{32} & m_{33}
	\end{bmatrix},
\end{equation}
where $m_{11}$, $m_{22}$, and $m_{33}$
denote the inertia coefficients associated
with the USV's surge, sway, and yaw motions,
respectively, while $m_{23}=m_{32}$
characterizes the inertial coupling
between sway and yaw.

The Coriolis and centripetal matrix $\mathbf C$ is given by
\begin{equation}
	\mathbf C =
	\begin{bmatrix}
		0 & 0 & c_{13}\\
		0 & 0 & c_{23}\\
		-c_{13} & -c_{23} & 0
	\end{bmatrix},
\end{equation}
where $c_{13}$ and $c_{23}$ characterize
the dynamic coupling effects between
the USV's surge and yaw motions and
between its sway and yaw motions,
respectively.

The hydrodynamic damping matrix
$\mathbf D$ is expressed as
\begin{equation}
	\mathbf D =
	\begin{bmatrix}
		d_{11} & 0 & 0\\
		0 & d_{22} & d_{23}\\
		0 & d_{32} & d_{33}
	\end{bmatrix},
\end{equation}
where $d_{11}$, $d_{22}$, and $d_{33}$
denote the damping coefficients in the
surge, sway, and yaw directions,
respectively, while $d_{23}$ and
$d_{32}$ describe the damping coupling
between sway and yaw.
All the above coefficients are assumed to be known constants under the adopted simplified USV dynamic model~\cite{10033669}.

\section{Wave-Induced Load Function}
\label{app:wave_load}

For completeness, the wave-induced load function
$\boldsymbol{\mathcal G}(\cdot)$ in \eqref{hailang}
can be expressed in terms of the steady
wave-induced forces and moment as~\cite{wicaksono2018wave}
\begin{equation}
	\boldsymbol{\mathcal G}
	\left(
	k_{0,n},\omega_{e,n},\chi_n,U_n,
	\boldsymbol{\Theta}_{\mathrm{usv}}
	\right)
	=
	[
		-\mathcal R_n,	\:\:	\mathcal Y_n,	\:\:
		\mathcal N_n
	]^T,
	\label{eq:wave_load_vector}
\end{equation}
where $\mathcal R_n$, $\mathcal Y_n$, and $\mathcal N_n$ denote the added resistance, steady sway force, and steady yaw moment, respectively, normalized by $\rho_w g a_n^2$.
The negative sign in the first component indicates that the added resistance acts opposite to the positive surge direction. Specifically, the normalized added resistance is given by
\begin{equation}
	\begin{aligned}
		\mathcal R_n
		&=
		\frac{1}{4\pi k_{0,n}}
		\int_{-\infty}^{\infty}
		W_n(q)
		\left(
		|C_n(q)|^2+|S_n(q)|^2
		\right)\\
		&\quad\times
		\frac{
			\Lambda_n(q)
			(q-k_{0,n}\cos\chi_n)
		}{
			\sqrt{\Lambda_n^2(q)-q^2}
		}
		\,dq,
	\end{aligned}
	\label{eq:wave_resistance}
\end{equation}
the steady sway force component is
\begin{equation}
	\begin{aligned}
		&\mathcal Y_n
		=
		-\frac{1}{4\pi k_{0,n}}
		\int_{-\infty}^{\infty}
		W_n(q)\Lambda_n(q)
		\Bigg[
		2\operatorname{Im}
		\left\{
		C_n(q)S_n^*(q)
		\right\}\\
		&\qquad-
		\left(
		|C_n(q)|^2+|S_n(q)|^2
		\right)
		\frac{
			k_{0,n}\sin\chi_n
		}{
			\sqrt{\Lambda_n^2(q)-q^2}
		}
		\Bigg]\,dq,
	\end{aligned}
	\label{eq:wave_sway}
\end{equation}
and the steady yaw moment component is
\begin{equation}
	\begin{aligned}
		&\mathcal N_n
		=
		\frac{1}{4\pi k_{0,n}}
		\int_{-\infty}^{\infty}
		W_n(q)\Lambda_n(q)\\
		&\quad\times
		\operatorname{Re}
		\left\{
		C_n'(q)S_n^*(q)
		-
		S_n'(q)C_n^*(q)
		\right\}
		\,dq\\
		&\quad-
		\frac{\sin\chi_n}{2}
		\operatorname{Re}
		\Bigg[
		H_n'(k_{0,n},\chi_n)\\
		&\qquad+
		\frac{1}{k_{0,n}}
		\left(
		\eta_n+
		\frac{k_{0,n}\cos\chi_n}{K_{U,n}}
		\right)
		H_n(k_{0,n},\chi_n)
		\Bigg].
	\end{aligned}
	\label{eq:wave_yaw}
\end{equation}

In the above expressions, $q$ denotes the integration wavenumber, while $C_n(q)$ and $S_n(q)$ denote the symmetric and antisymmetric Kochin functions, respectively.
These functions characterize the wave disturbances induced by the USV and depend
on its hull geometry and hydrodynamic parameters $\boldsymbol{\Theta}_{\mathrm{usv}}$.
The function $H_n(k_{0,n},\chi_n)$ denotes the directional Kochin function evaluated at the incident wave conditions.
The wavenumber $\Lambda_n(q)$ associated with the frequency $\omega_{e,n}+qU_n$
is defined as $\Lambda_n(q)=(\omega_{e,n}+qU_n)^2/g$.
The dimensionless parameter $\eta_n=U_n\omega_{e,n}/g$ characterizes the effect of the USV forward speed on the wave encounter, while $K_{U,n}=g/U_n^2$ denotes the speed-dependent wavenumber scale.
Finally, the function $W_n(q)=\operatorname{sgn}(\omega_{e,n}+qU_n) \mathbf 1_{\{\Lambda_n^2(q)\geq q^2\}}$ selects the propagating wave components and accounts for the sign of the corresponding frequency, where $\operatorname{sgn}(\cdot)$ and $\mathbf 1_{\{\cdot\}}$ denote the sign and indicator functions, respectively.

\bibliographystyle{IEEEtran}
\bibliography{reference}

\end{document}